\documentclass [a4paper,11pt]{article}
\usepackage[english]{babel}
\usepackage{graphicx,amsmath,amsfonts,amssymb,slashed,upgreek,caption,amscd,cite,cancel}

\newcommand{\be}{\begin{equation}}
\newcommand{\ee}{\end{equation}}
\newcommand{\beq}{\begin{equation}}
\newcommand{\eeq}{\end{equation}}
\newcommand{\bea}{\begin{eqnarray}}
\newcommand{\eea}{\end{eqnarray}}

\newcommand{\A}{{\cal A}}

\newcommand{\R}{{}^{(3)}\!R}
\newcommand{\G}{{}^{(3)}\!G}
\newcommand{\nd}{{\dot n}}
\newcommand{\g}{\gamma}
\newcommand{\B}{{\cal B}}
\newcommand{\C}{{\cal C}}

\newcommand{\M}{{\cal M}}
\newcommand{\Y}{{\cal Y}}
\newcommand{\sR}{{\cal R}}
\newcommand{\SK}{{\cal S}}
\newcommand{\F}{{\cal F}}
\newcommand{\Z}{{\cal Z}}

\usepackage[stable]{footmisc}
\def\d{\delta}

\def\MM{M_{*}}
\newcommand{\mfs}{{m}_4^2}
\newcommand{\tmfs}{{\tilde m}_4^2}

\begin{document}

\begin{center}
\Large{\textbf{Essential Building Blocks of Dark Energy}} \\[1cm] 
 
\large{Jerome Gleyzes$^{\rm a}$,  David Langlois$^{\rm b}$, Federico Piazza$^{\rm b,c}$ and Filippo Vernizzi$^{\rm a}$}
\\[0.5cm]

\small{
\textit{$^{\rm a}$ CEA, IPhT, 91191 Gif-sur-Yvette c\'edex, France CNRS, \\ URA-2306, 91191 Gif-sur-Yvette c\'edex, France}}

\vspace{.2cm}

\small{
\textit{$^{\rm b}$  APC, (CNRS-Universit\'e Paris 7), 10 rue Alice Domon et L\'eonie Duquet, 75205 Paris, France \\ 
}}
\vspace{.2cm}

\small{
\textit{$^{\rm c}$ PCCP, 10 rue Alice Domon et L\'eonie Duquet, 75205 Paris, France \\ 
}}

\end{center}

\vspace{2cm}

\begin{abstract}
We propose a minimal description of  single field dark energy/modified gravity within  the effective field theory formalism for cosmological perturbations, which encompasses most  existing models.  We start from a generic   Lagrangian given as  an arbitrary function
of the lapse  and of the extrinsic and intrinsic curvature tensors of the time hypersurfaces in unitary gauge, i.e. choosing as time slicing the uniform scalar field hypersurfaces. Focusing on linear perturbations, we identify seven Lagrangian operators that lead  to  equations of motion containing at most two (space or time) derivatives, the  background evolution being determined by the time dependent coefficients of only three of these operators.
We then establish a dictionary that translates any existing or future model whose Lagrangian can be written in the above form  into our parametrized framework.  
As an illustration,  we study 
 Horndeski's---or generalized Galileon---theories and show that they can be described, up to linear order, by only six  of the seven operators mentioned above. 
 This implies, remarkably, that the dynamics of linear perturbations can be more general than that of Horndeski while remaining second order.
 Finally, in order to make the link with observations, we provide the entire set of linear perturbation equations in Newtonian gauge, the effective Newton constant in the quasi-static approximation and  the ratio of the two gravitational potentials, in terms of the time-dependent coefficients of our  Lagrangian.

\end{abstract}

\newpage 
\tableofcontents

\vspace{.5cm}

\section{Introduction}

Dark energy has now become a generic name that includes a huge number of models 
trying to account for the present cosmic acceleration~\cite{costas,lucashin}. Given their proliferation, the confrontation  of such models with present and future cosmological data would be greatly facilitated by  an effective approach that can mediate between  observational data and  theory. 
Ideally, such a phenomenological approach would provide  an effective parameterisation that minimizes the number of free functions and deals directly with the relevant low-energy degrees of freedom, which in our context are the cosmological perturbations (together with the background evolution). A precise dictionary rephrasing the various models into this common language would then simplify the confrontation with the data
 and point out possible degeneracies between different theories. Within its unifying picture, this effective approach should have the extra virtue of  stimulating  theorists  to study previously unexplored regions of the parameter space which could lead to interesting new  models or, conversely,  to better understanding why  certain regions might be forbidden.

{A few steps in  this direction have been undertaken recently. The so-called effective field theory (EFT) of cosmological perturbations is a powerful tool that allows to deal directly with the relevant low-energy degrees of freedom of the problem at hand.} 
{Such} an approach was proposed and intensively used  for inflation~\cite{EFT1,EFT2}, in particular to characterize high-energy corrections to slow-roll models and {to} predict high-order correlation functions  (see e.g.~\cite{Senatore:2009gt,Creminelli:2010qf}). {The EFT of inflation}  has now become a standard way of parametrising  primordial non-Gaussianity and  was used, for instance, in the interpretation of the most recent  WMAP \cite{Bennett:2012fp} and Planck \cite{Ade:2013ydc} data.
This approach has also been applied to dark energy,  first in the minimally-coupled case~\cite{Creminelli:2008wc} where it was proven a useful tool to study the stability in full generality and, for models with vanishing sound speed, the clustering of dark energy  down to very nonlinear scales~\cite{Creminelli:2009mu}.

More recently, the EFT formalism has been extended to dark energy with non-minimal couplings~\cite{GPV,BFPW},  providing a unifying theoretical framework for practically   all single-field dark energy and modified gravity models.\footnote{An alternative formulation of a background independent effective approach to dark energy and modified gravity was given in \cite{Battye:2012eu}. A covariant EFT of cosmological acceleration was developed in \cite{wein-eft} for inflation and generalized to the case of dark energy in \cite{PZW,BF}---see discussion in \cite{GPV} for  a comparison {between} the latter approach {and} the one advocated here. For a different unifying framework to cosmological perturbations for dark energy and modified gravity see, for instance, \cite{Baker:2012zs}.} 
This approach relies {on}  two basic steps \cite{GPV}:
{\it   a)} assume  the weak equivalence principle and therefore the existence of a metric $g_{\mu \nu}$ universally coupled to all matter fields (it  is straightforward to  relax this assumption, but at the price of complicating the formalism);
{\it  b)} write {the {\it unitary gauge} action, i.e.~the most general gravitational action for such a metric compatible with {the (unbroken)} spatial diffeomorphism invariance on hypersurfaces of constant dark energy field. }

In \cite{GPV} it was argued that the EFT of dark energy has  all the virtues advocated at the beginning of this section. The goal of this article is to provide a systematic procedure to translate an arbitrary dark energy model into the EFT language, as well as to establish {a firm} minimal setting of Lagrangian operators within this framework. 
In particular, here we focus our attention on  the  operators of the unitary gauge action that lead to at most {two} derivatives in the equations of motion for linear  perturbations.\footnote{This is sufficient to ensure that we only have  a single propagating degree of freedom. Note, however, that higher time derivatives do not  lead to higher degrees of freedom if they can be treated perturbatively, i.e.~evaluating them using the lower order equations of motion \cite{Simon:1990ic}.} This minimal set of operators encompasses most of the theoretical models of dark energy and/or modified gravity discussed in the current literature. 

The key ingredient of our derivation is a $3+1$ decomposition {\it \`a la} ADM, where  time slicings coincide with the uniform scalar field hypersurfaces. With this time choice, the dynamics of the underlying degree of freedom is embodied in the dynamics of the 3-dimensional metric.
In Sec.~\ref{sec_2} we consider  a generic Lagrangian given as an arbitrary function of the lapse $N \equiv 1/\sqrt{-g^{00}}$ and of the 3-dimensional metric $h_{\mu \nu} \equiv g_{\mu \nu} +n_\mu n_\nu$, where $n^\mu$ is the unit vector perpendicular to constant time hypersurfaces, more specifically of its extrinsic and intrinsic curvature tensors, respectively $K^{\ \mu}_{\nu} \equiv h_{\nu}^{\ \rho} \nabla_\rho n^\mu$ and $\R^\mu_{\ \nu}$,
\beq
\label{intro_action}
S=\int d^4 x \sqrt{-g}\,  L(N,  K^\mu_{\ \mu}, K_{\mu \nu} K^{\mu \nu}, \R, \R_{\mu \nu} \R^{\mu \nu}, \ldots ;t) \;.
\eeq
In our construction we  include combinations of these 3-dimensional objects {\it without} taking their derivatives. This automatically prevents the appearance of higher (more than two) {\it time} derivatives in the equations of motion.
However, it is not enough to also remove  higher {\it spatial} derivatives.
By expanding this Lagrangian up to quadratic order in the cosmological perturbations and making use of an ADM analysis in unitary gauge (see for instance \cite{EFT1,malda,Boubekeur:2008kn}) we  obtain specific conditions that ensure the absence of higher spatial derivatives in Sec.~\ref{sec:ADM}. 

Moreover, we also show how the parameters in front of the standard EFT operators of \cite{GPV} can be expressed in terms of the time-dependent coefficients of the expansion of \eqref{intro_action}. Since the action of most  of the existing theoretical models can be written as eq.~\eqref{intro_action}, this can be used to derive a dictionary between theoretical models and our EFT language. 
As an illustration, in Sec.~\ref{section_Galileons} we explicitly derive this dictionary for the most general scalar field theory leading to at most second order equations of motion, i.e.~the Horndeski theory \cite{horndeski} (see also\cite{fab-four}), recently rediscovered in the context of the so-called Galileon field \cite{NRT,Deffayet:2009wt} under the name of ``generalized Galileons'' \cite{Deffayet:2009mn,DGSZ}.

Let us summarize here the main results of Secs.~\ref{sec_2} and \ref{section_Galileons}:
\begin{itemize}
\item The most general EFT action,  up to quadratic order, for single-field dark energy, in the  Jordan frame, leading to at most second-order  equations of motion for {\it linear} perturbations can be written as 
\be
\begin{split}
\label{total_action}
S =& \int \! d^4x \sqrt{-g} \left[\, \frac{\MM^2}{2} f(t) R - \Lambda(t) - c(t) g^{00}  + \, \frac{M_2^4(t)}{2} (\delta g^{00})^2\, -\, \frac{m_3^3(t)}{2} \, \delta K \delta g^{00} \, 
 \right. \\[1.2mm]
 &  - \left. \,  m_4^2(t)\left(\delta K^2 - \d K^\mu_{ \ \nu} \, \d K^\nu_{ \ \mu} \right) \, +\, \frac{\tilde m_4^2(t)}{2} \, \R \, \delta g^{00}  \right] \;,
\end{split}
\ee
where $\delta g^{00} \equiv g^{00} +1$, $\delta K_{\mu \nu} \equiv K_{\mu \nu} - H h_{\mu \nu}$, $K \equiv K^{\mu}_{\ \mu}$ and we have assumed a flat Universe so that $\R^{\mu}_{\ \nu}$ vanishes on the background\footnote{The case of non-vanishing spatial curvature is commented on in footnote~\ref{footnote}.}. This action describes the propagation of one scalar degree of freedom with dispersion relation $\omega^2 = c_s^2 k^2$, where $c_s$ is the sound speed of fluctuations given by eq.~\eqref{sound_speed} with the relations \eqref{zeta_Lagrangian_EFT}. Stability (absence of ghosts) is ensured by the positivity of the time kinetic term given in \eqref{zeta_Lagrangian_EFT}. The particular combination appearing in the operator proportional to $m_4^2$ is such that it does not lead to higher-order spatial derivatives.  One can check that also the combination 
\be\R^\mu_{\,\nu} \, \d K_\mu^{\nu}  - \frac12  \R \, \delta K  
\ee 
does not generate higher derivatives. However, this operator is not explicitly included in eq.~\eqref{total_action} because it can be reexpressed in terms of the others (see App.~\ref{app:KR}). 

\item In the particular case where $m_4^2=\tilde m_4^2$, the above action is equivalent to the {\it linearized} Horndeski's theory/generalized Galileons and the explicit dictionary between generalized Galileons and this action is given in App.~\ref{app:Galileon_dic}.  This implies that the dynamics of linear scalar perturbations of action \eqref{total_action} is more general than that of Horndeski, while remaining  second order in time and space derivatives.

\item Expanding the Lagrangian \eqref{intro_action} up to quadratic order we also find  three operators that lead to higher order space---but not time---derivatives. These are 
\be
\label{hsd}
S_{\rm h.s.d.} = \int \! d^4x \sqrt{-g} \left[ - \,  \bar m_4^2(t)\, \delta K^2 \, +\, \frac{\bar m_5(t)}{2} \, \R \, \delta K \, + \frac{\bar \lambda(t)}{2} \R^2 \right] \; .
\ee
When one of these operators is present in the action the dispersion relation of the propagating mode receives corrections at large momenta, $\omega^2 = c_s^2 k^2 + k^4/M^2$, where $M$ is a mass scale. These corrections may become important in the limit of vanishing sound speed, such as in the model of the Ghost Condensate~\cite{ghost} or for deformations of this particular limit \cite{EFT1,Creminelli:2008wc}.

\end{itemize}

Once a Lagrangian describing matter has been  included, the action \eqref{total_action} can be used as a benchmark for the study of physical signatures of dark energy/modified gravity  in the linear regime.  In this context, cosmological perturbations are usually discussed in Newtonian gauge, which is the one that we  employ in Sec.~\ref{sec:obs}. In order to do that, in Sec.~\ref{sec:pert_eqs} we restore the covariance via the Stueckelberg trick \cite{ghost,EFT1,EFT2} and we vary this action with respect to all the scalar dynamical degrees of freedom. This allows to derive  Einstein's equations and the evolution equation for the fluctuations of the scalar field responsible of the acceleration, recovering and generalizing previous results~\cite{Creminelli:2008wc,GPV}. 
Interestingly, in the Newtonian gauge  Einstein's equations and the scalar equation of motion  contain higher order derivatives when $m_4^2\neq\tilde m_4^2$, even if the dynamical equation for the true degree of freedom is only second order.
Finally, in Sec.~\ref{sec:mod_grav} we use these equations to derive the effective Newton constant and the ratio between the two gravitational potentials.

The phenomenology of the operators appearing in action \eqref{total_action} was also studied  in \cite{BFPW}. In this reference it was indeed mentioned that these operators are sufficient to describe linear perturbations of Horndeski's theories. However, no proof of this statement was given nor the particular combination in which such operators appear shown.

\section{General Lagrangian in unitary gauge}
\label{sec_2}

In the presence of a scalar field $\phi$ with a non-vanishing timelike gradient, the so-called unitary gauge 
corresponds to  a choice of  time slicing  where the constant time hypersurfaces are  uniform $\phi$ hypersurfaces.
The use of unitary gauge accomplishes two main objectives. First, as explained at length in Refs.~\cite{ghost,EFT1,EFT2}, it makes it straightforward to write a generic Lagrangian for cosmological perturbations. Since the dynamics of the scalar field has been ``eaten" by the metric, the most generic Lagrangian is simply that for the {\it metric perturbations} around a FLRW solution, compatible with the unbroken symmetry of 3-dimensional diffeomorphisms. 

Second, the 3+1 splitting in unitary gauge easily allows to keep the number of time derivatives under control, while considering higher and higher space derivatives. 
Therefore, the unitary gauge is helpful to systematically explore  the space of higher spatial derivative theories by considering geometric invariants on the $\phi = constant$ hypersurfaces. In practice, we will use the metric in the ADM form  
\be
\label{ADM}
ds^2=-N^2 dt^2 +{h}_{ij}\left(dx^i + N^i dt\right)\left(dx^j + N^j dt\right) \, ,
\ee
where the 3-dimensional metric $h_{ij}$ is used to lower and raise latin indices $i,j,\dots = 1,2,3$.
Since 3-dimensional diffeomorphism invariance is preserved in unitary gauge, it is natural to write  operators (with up to two spatial derivatives per field)  in terms of the extrinsic and intrinsic curvatures $K_{\mu \nu}$ and $\R_{\mu \nu}$ and their possible contractions. The Lagrangian  is also an explicit function of the lapse function $N$ in general.

Therefore, in the following, we  consider a general action of the form
\beq
\label{start_point}
S=\int d^4 x \sqrt{-g}\,  L(N,  K, \SK, \sR,\Z;t) \;, 
\eeq
where the Lagrangian  $L$ is  an arbitrary function of $N$ and of the  following four scalar quantities constructed by contracting the extrinsic and intrinsic curvature tensors: 
\be
K\equiv K_{\ \mu}^\mu\;, \qquad \sR \equiv \R\equiv\R_{\ \mu}^\mu \;, \qquad \SK\equiv K_{\mu\nu}K^{\mu\nu}, \quad \Z\equiv \R_{\mu\nu}\R^{\mu\nu}\;.
\ee
Although one should also allow, in principle, for a dependence on  $ \Y\equiv {}^{(3)}\!R_{\mu\nu}K^{\mu\nu}$, we have preferred not to include it explicitly in the main body, for simplicity. As shown  in   App.~\ref{app:KR}, 
  this extra dependence leads to a quadratic Lagrangian of the same form as that found later in this section, with slightly modified   coefficients.  
Indeed, since $\Y$ is equivalent to  $H \sR$ at linear order, the  quadratic terms in the expansion of the Lagrangian induced by its dependence on $\Y$  are analogous to those induced by its dependence on $\sR$. As for the linear term, one can use 
 the equality
\be
 \lambda(t) {}^{(3)}\!R_{\mu\nu}K^{\mu\nu}  \ =\  \frac{\lambda(t)}{2} {}^{(3)}\!R\; K \ + \ \frac{\dot \lambda(t)}{2 N} \; \R \ + {\rm boundary} \ {\rm terms}\;,
\ee
which is also shown  in App.~\ref{app:KR}.

Moreover, one could also consider scalars that are combinations of three or more tensors, like $K^{\lambda}_{\ \mu} K^\mu_{\ \nu} K_{\ \lambda}^{\nu}$, but it is easy to show that also  in this case they can be re-expressed in terms of the above combinations, plus corrections which are at least cubic in the perturbations. We will show this explicitly for the extended Galileon in the next section. 
Finally, one could take quadratic combinations of the Riemann tensor such as $\R_{\mu \nu \rho \sigma} \R^{\mu \nu \rho \sigma}$. However, in three dimensions the Riemann tensor can be expressed in terms of the Ricci scalar and tensor.\footnote{This can be done using the relation  \be
\R_{\mu \nu \rho \sigma} = \R_{\mu \rho} h_{\nu \sigma} - \R_{\nu \rho} h_{\mu \sigma} - \R_{\mu \sigma} g_{\nu \rho} + \R_{\nu \sigma}h_{\mu \rho} -\frac12 \R (h_{\mu \rho} h_{\nu \sigma} - h_{\mu \sigma} h_{\nu \rho}) \;.
\ee} Thus, at quadratic order in the perturbations, 
the action above seems to exhaust  all the possibilities compatible with our requirements.

In order to explicitly write  the expansion of the action \eqref{start_point} up to second order in the perturbations, it is useful to define the tensors
\beq
\label{dK}
\d K\equiv K-3H,\qquad  \d K_{\mu\nu}\equiv K_{\mu\nu}-H h_{\mu\nu}\,,
\eeq
which vanish on the background, 
and to use the decompositions
\be
\label{dS}
\SK=3H^2+\delta \SK, \qquad \delta \SK \equiv 2H \d K +\d K^\mu_{ \ \nu} \d K^\nu_{ \  \mu}  \;.
\ee
The quantities 
$\sR$ and $\Z$ vanish on the background and are therefore already perturbative ($\Z$ is even a second order quantity).

The expansion of the Lagrangian  up to second order in the perturbations yields, after discarding irrelevant boundary terms, the expression
\be
\label{lag}
\begin{split}
L(N,K,\SK,\sR,\Z)&= \bar{L} - \dot \F -3H \F +( \dot \F + L_N) \, \d N+L_\sR\, \d \sR 
\cr
&+ \frac{\A}{2} \, \d K^2 +L_\SK\, \d K^\mu_{ \ \nu} \d K^\nu_{ \ \mu}+\left( \frac12 L_{NN} - \dot \F \right)\d N^2\cr
&+\frac12  L_{\sR\sR} \, \d \sR^2
+\B \, \d K \d N +\C \, \d K \d \sR+ L_{N\sR} \, \d N \d \sR + L_\Z \d \Z +{\cal O}(3)\,,
\end{split}
\ee
where we have introduced the following notations for some combinations of the partial derivatives of the Lagrangian (denoting $L_N\equiv \partial L/\partial N$, etc.), to make this expression more compact:
\be 
\begin{split}
 \F & \equiv 2H L_\SK+L_K\,, \\
\A &\equiv 4H^2 L_{\SK \SK }+4H L_{\SK K}+ L_{KK}\,,\\  
\B & \equiv 2H L_{\SK N}+ L_{KN}\,, \\
 \C& \equiv 2H L_{\SK \sR}+ L_{K\sR}  \;.
\end{split}
\ee
The first term, $\bar{L}$, is the homogeneous Lagrangian and all partial derivatives of $L$ that appear in the above expression are evaluated on the homogeneous background, i.e. for $\bar{N}=1$, $\bar{\SK }=3H^2$, $\bar{K}=3H$, 
$\bar{\sR}=0$ and $\bar{\Z}=0$. 
Note that, in order to obtain the expression (\ref{lag}),  we have  rewritten the term linear in $\delta K $ as 
\be
\F \delta K = \F (K - 3H ) \;, 
\ee
and integrated it by parts using $K=\nabla_\mu n^\mu$,
\beq
\int d^4x \sqrt{-g} \, \F K=-\int d^4x \sqrt{-g} \, n^\mu \nabla_\mu {\cal F}=-\int d^4x \sqrt{-g} \frac{\dot {\cal F}}{N}\,,
\eeq
where $n^\mu$ is the unit vector  orthogonal to  constant time hypersurfaces and,  in unitary gauge,  has time component $n^0 = 1/N$.

\subsection{Background equations}
\label{sec:be}
For the background we assume a flat homogeneous  FLRW metric, written in the form
\beq
ds^2=-N^2(t) dt^2 +a^2(t) \delta_{ij} dx^i dx^j\,.
\eeq
In this case $K=3H /N$ and $\SK =  3 H^2/N^2$, where $H \equiv \dot a/a$ is the Hubble rate.
Note that it is crucial to explicitly keep the lapse function $N$, because the first Friedmann equation is the constraint  arising from the invariance under time reparametrization. Linear variation of the homogeneous action $S_0$ with respect to the lapse $N$ and the scale factor $a$ yields
\beq
\d S_0=  \int dt d^3x\left[ a^3\left(\bar{L}+ L_N-3H \F \right)\d N+3 a^2\left(\bar{L} -3H\F -\dot \F\right) \delta a\right] \;,
\eeq
where we have used $\sqrt{-g} = a^3 N$.
Then the Friedmann equations are directly given by\footnote{We have not included explicitly the matter in the Friedmann equations, but it is straightforward to do so.} 
\beq
\label{beom1}
3H\F- \bar{L}-L_N=0,
\eeq
which depends on first time derivatives at most, and
\beq
\label{beom2}
\dot \F+3H \F-\bar{L}=0\,,
\eeq
which determines the dynamics of the scale factor.

As expected, by using the above equations one can check that the first order of the total Lagrangian ${\cal L}\equiv \sqrt{-g}\,  L$ vanishes. Indeed, using $\sqrt{-g}=\sqrt{h} \, N$, where $h$ is the determinant of the  3-dimensional metric $h_{ij}$ in the ADM decomposition, one easily finds
\bea
{\cal L}_1=\left(\bar{L}-3H\F-\dot \F\right) \d\sqrt{h}+a^3(L_N+\bar{L}-3H\F)\d N +a^3 L_\sR \, \d \sR\,,
\eea
where the last term is a total derivative and can be ignored. 

\subsection{Perturbations in the ADM formalism}
\label{sec:ADM}

In this sub-section we perform the analysis of the perturbations in unitary gauge and by using the ADM form of the metric, eq.~\eqref{ADM}. For the action at second order, we will only need to take into account  the perturbations of $\sqrt{-g} $ at first order, $\delta \sqrt{-g} = \delta \sqrt{h} + a^3 \delta N$, because the second order one multiplies the LHS of eq.~\eqref{beom2}. Thus, the  quadratic Lagrangian for perturbations is given by
\be
\label{Lquad}
\begin{split}
{\cal L}_2&=  \d  \sqrt{h} \left[ (\dot \F+L_N )\d N+ L_\sR\, \d \sR\right]
\cr
&
+ a^3\left[\left(L_N+\frac12 L_{NN}\right)\d N^2+L_\sR \d_2\sR +\frac12\A \, \d K^2 
+\B \, \d K \d N
+
\C\, \d  K \d \sR
\right.
\cr
& \left.
+ L_\SK \, \d K^\mu_\nu \, \d K^\nu_\mu
+L_\Z \,\d \sR^\mu_\nu \, \d \sR^\nu_\mu
+ \frac12  L_{\sR\sR} \; \d \sR^2
+ (L_\sR + L_{N\sR}) \, \d N \d\sR
\right] \;, 
\end{split}
\ee
where $\d_2\sR$ denotes the expansion of $\sR$ at second order in the perturbations.

In the ADM decomposition \eqref{ADM} the
only relevant 
 components of the extrinsic curvature tensor are given by
\beq
\label{KijADM}
K_{ij}=\frac{1}{2 N}(\dot{h}_{ij} - \nabla_iN_j-\nabla_jN_i) \;, \\
\ee
where $\nabla_i$ stands for  the covariant derivative associated with the 3-dimensional metric $h_{ij}$. For explicit calculations in unitary gauge we choose to describe scalar perturbations of the spatial metric in terms of $\zeta$ \cite{malda},
\beq
{h}_{ij}=a^2(t) e^{2\zeta}\delta_{ij}\,.
\eeq
(We consider the tensor modes separately in App.~\ref{app:tensor}.)
Thus, the perturbations of the quantities used above are given by
\be
\label{dhdK}
\d\sqrt{h} = 3 a^3 \zeta\,, \qquad  \d K^i_{\ j}=\left(\dot\zeta-H\d N\right)\d^i_j-\frac{1}{a^2}\d^{ik}\partial_{(k}N_{j)} \;,
\ee
and
\be \label{RRR}
\d\sR_{ij} =  - \delta_{ij} \partial^2 \zeta -  \partial_i \partial_j \zeta \;, \qquad \d_2 \sR=  -\frac{2}{a^2}\left[(\partial\zeta)^2-4\zeta\partial^2\zeta\right]\,.
\ee
Note that in this section $\partial$ stands for a spatial derivative and $\partial ^2 \equiv \partial_i \partial ^i$. 
By using the above expressions, the variation of ${\cal L}_2$ 
with respect to $\d N$ yields the Hamiltonian constraint, which reads
\bea
&&\left[L_{NN}+2L_N+3H\left(3H\A+2H L_\SK -2\B\right)\right]\d N +3\left(\B-3H\A-2H L_\SK\right)\dot\zeta +3(L_N+\dot\F)\zeta
\cr
&& - \left(\B - 3H\A - 2H L_\SK  \right)\frac{\partial^2\psi}{a^2}-4\left(L_\sR+L_{N\sR}-3H\C\right)\frac{\partial^2\zeta}{a^2}=0\,.
\eea
By varying ${\cal L}_2$ with respect to the shift 
\be
\label{shift}
N_i \equiv \partial_i\psi\;, 
\ee
one obtains the momentum constraint, which implies 
\beq
\label{constraint}
-\left(\B - 3H\A - 2H L_\SK \right)\d N+\left(\A+2L_\SK \right)\frac{\partial^2\psi}{a^2}=\left(3\A+2L_\SK \right)\dot\zeta- 4 \C \frac{\partial^2\zeta}{a^2}\,.
\eeq
By combining the two constraints, one can express both $\delta N$ and $\partial^2\psi$ as functions of $\zeta$ and its derivatives and then substitute in the action to write it only in terms of $\zeta$ and its derivatives.  In general, a term proportional to $(\partial^2\zeta)^2$ will remain. Here, in order to single out the lowest derivatives operators first,  we want to find  conditions under which this term disappears. If one considers the second order action before the substitution of the constraints, one finds the following terms
\beq
\frac{1}{a}\left[\frac12\left(\A+2L_\SK\right) (\partial^2\psi)^2+ 4 \C \; \partial^2\psi \, \partial^2\zeta+2\left(4 L_{\sR \sR}+3L_\Z\right) (\partial^2\zeta)^2 \right]\,.
\eeq
Taking into account the momentum constraint (\ref{constraint}), one immediately  sees that imposing the three conditions\footnote{Note that these  conditions are only sufficient. A more general analysis can be performed by explicitly requiring  that  the coefficient of $(\partial^2\zeta)^2$ vanishes once the two constraints have been solved. However, this leads to a very complicated equation involving many of the coefficients of the quadratic Lagrangian and it is not clear whether one can find physically relevant solutions that  evade \eqref{conds}.}
\beq
\label{conds}
\A+2L_\SK =0\;, \qquad \C =0 \; ,\qquad 4L_{\sR \sR}+3L_\Z=0\; ,
\eeq
implies  the elimination of  the term proportional to $(\partial^2\zeta)^2$ in the final action and the absence of higher derivatives in the equation of motion for $\zeta$. As we will see in the next section, all generalized Galileon models satisfy the three conditions~\eqref{conds}.

When~\eqref{conds} are satisfied, the momentum constraint reduces to  
\beq 
\d N\, =\, {\cal D} \, \dot\zeta\; , \qquad {\cal D}\equiv\frac{4L_\SK }{\B+4HL_\SK } \; .
\eeq
By direct substitution into ${\cal L}_2$ and after an integration by parts to get rid of the term $\dot\zeta\partial^2\zeta$ (note that the $\zeta\dot\zeta$ term vanishes because of the background equations of motion), we finally get the following Lagrangian for $\zeta$:
\be
\label{quadratic_action_zeta}
{\cal L}_2 = \frac{a^3}2 \left[ {\cal L}_{\dot \zeta \dot \zeta}  \,  \dot\zeta^2
+ {\cal L}_{\partial_i \zeta \partial_i \zeta} \, \frac{(\partial_i \zeta)^2}{a^2}\right],
\ee
with
\be
\label{zeta_Lagrangian}
\begin{split}
{\cal L}_{\dot \zeta \dot \zeta} &\equiv  2 \left( \frac12 L_{NN}+L_N - 3 H \B - 6 H^2 L_\SK \right){\cal D}^2+12L_\SK \;, \\
{\cal L}_{\partial_i \zeta \partial_i \zeta} &\equiv  4\left[ L_\sR - \frac{1}{a}\frac{d}{dt}(a\M) \right] \;, \qquad \M\equiv  {\cal D} (L_\sR  + L_{N\sR} ) \,.
\end{split}
\ee
Classical and quantum stability (absence of ghosts) requires that the time kinetic energy is positive (see, e.g.~\cite{EFT1,Creminelli:2008wc}),
\be
\label{stability}
{\cal L}_{\dot \zeta \dot \zeta}>0\;.
\ee
The sound speed (squared) of fluctuations can be simply computed by taking the ratio
\be
\label{sound_speed}
c_s^2 = -\frac{ {\cal L}_{\partial_i \zeta \partial_i \zeta} }{ {\cal L}_{\dot \zeta \dot \zeta} } \;.
\ee

\subsection{The EFT language} \label{sec_2.3}

We are now going to express the conditions on the absence of higher derivatives in terms of the coefficients of the action of the EFT formalism of Refs.~\cite{GPV,BFPW}. The action up to quadratic order in the perturbations can be written as
\be
\begin{split}
\label{total_action_2}
S =& \int \! d^4x \sqrt{-g} \left[\, \frac{\MM^2}{2} f R - \Lambda  - c  g^{00} +  \frac{M_2^4 }{2} (\delta g^{00})^2 - \frac{\bar m_1^3 }{2}  \delta K \delta g^{00} 
 \right. \\[1.2mm]
 &  - \left.  \frac{\bar M_2^2 }{2}  \delta K^2 - \frac{\bar M_3^2 }{2}\delta K^{\mu}_{\ \nu} \, \delta K_{\mu}^{\  \nu} + \frac{\mu_1^2 }{2}  \R  \delta g^{00}   + \frac{\bar m_5 }{2}  \R  \delta K + \frac{\lambda_1}{2} \R^2 + \frac{\lambda_2}{2} \R^{\mu}_{\ \nu } \R_{\mu }^{\ \nu} \right] ,
\end{split}
\ee
where $R$ in the first term inside the bracket is the four-dimensional Ricci scalar. Note that, in order to make the comparison with the previous subsection simpler, we have found more convenient to use the 3-dimensional Ricci scalar and tensor in the quadratic terms, instead of the four-dimensional ones used in Ref.~\cite{GPV}, since the link with the ADM decomposition is then transparent.

Let us first discuss how the background equations  \eqref{beom1} and \eqref{beom2} translate in this language. In action \eqref{total_action_2} we have used the time-time component of the inverse metric  $g^{00}$ and its perturbation in the expansion of quadratic and higher order operators, as it is customary in the EFT formalism.
However, in the previous subsections it was more convenient to work directly with the lapse function $ N$,  related to $g^{00}$ by
\beq
\label{g00_N}
g^{00}=-\frac{1}{N^2} \;.
\eeq
Only the first three terms in  brackets in eq.~\eqref{total_action_2} contribute to $\bar L$, $L_N$ and $\F$, and thus to the background equations of motion.
Using eq.~\eqref{g00_N} and employing the decomposition of the four-dimensional curvature scalar, 
\be \label{RR}
R = \R + K_{\mu \nu} K^{\mu \nu} -  K^2 + 2 \nabla_\nu(n^\nu \nabla_\mu n^\mu-n^\mu \nabla_\mu n^\nu) \;,
\ee
after an integration by parts in the action we can rewrite these terms as
\be
\label{L0_EFT}
L_0 = \frac{\MM^2}{2} \left( f  \sR + f {\cal S} - fK^2 -2 \dot f \frac{K}{N} \right) - \Lambda + \frac{c}{N^2} \;,
\ee
(we remind the reader that $\sR \equiv \R$). By expanding at linear order in $\delta N$, integrating by parts the terms linear in $K$, we can match the background and linear terms of this action with the first line of eq.~\eqref{lag}, which yields
\be
 \label{barL}
\begin{split}
\bar L - \dot \F - 3 H \F & =  \MM^2 ( 3 f H^2 + 2f \dot H +2 \dot f H + \ddot f )  + c - \Lambda \;,\\
\dot \F + L_N  &=  \MM^2 (  \dot f H-  2 f \dot H  - \ddot f ) -2 c \;.
\end{split}
\ee
From  these two relations and using the background equations of motion  \eqref{beom1} and \eqref{beom2} one finds that  $c$ and $\Lambda$ are given by 
\begin{align}
c+ \Lambda &= 3 \MM^2 \left(f H^2 + \dot fH \right) \;, \label{cccc} \\ 
\Lambda - c &= \MM^2 \left(2 f \dot H + 3 f H^2 + 2 \dot f H + \ddot f \right)\; . \label{lambdalambda}
\end{align}
This coincides with what was found in Ref.~\cite{GPV} in the absence of matter.

To discuss linear perturbations we only need the second-order expansion of the action \eqref{total_action_2}. 
By rewriting the first three terms  as in eq.~\eqref{L0_EFT}, expressing $g^{00}$ in terms of $N$ and using the definitions \eqref{dK} and \eqref{dS}, one immediately sees that the EFT action  is of the form \eqref{start_point}. One can thus use  the second-order expansion of the Lagrangian \eqref{Lquad} with  the following dictionary:
\be
\begin{split}
\label{relations}
L_\sR & = \frac12 \MM^2 f \;, \\
\frac12 L_{NN} + L_N &= c+ 2 M_2^4\;,\\  
\A & =  - \MM^2 f - \bar M_2^2  \;, \\ 
\B &= \dot f \MM^2 - \bar m_1^3 \;, \\ 
\C &= \frac{\bar m_5}{2}\;, \\ 
L_{{\cal S}} & = \frac12 \left(\MM^2 f - \bar M_3^2 \right)\;, \\
L_{\Z} &= \frac{\lambda_2}{2}\;, \\
L_{N \sR} &= \mu_1^2 \;, \\ 
L_{\sR \sR} & = \lambda_1 \; ,
\end{split}
\ee
which is completed  with eq.~\eqref{barL}. 

With these relations, the conditions for the absence of higher derivatives, eq.~\eqref{conds}, can be written in the EFT of dark energy language. They read:
\be
\label{dK_noder}
 \bar M_2^2 + \bar M_3^2 = 0\;, \qquad 
 \bar m_5 =0 \;, \qquad 
 4 \lambda_1 + \frac32 \lambda_2 = 0 \;.
\ee
These conditions are straightforward to verify. Using eqs.~\eqref{dhdK} and \eqref{shift}, $ \delta K^2$ contains a higher derivative term,  $(\partial^2 \psi)^2$, while $ \delta K^\mu_{\ \nu} \d K_\mu^{\ \nu}$  contains $(\partial_i \partial_j \psi)^2$. However, when the first condition in eq.~\eqref{dK_noder} is satisfied  the combination of higher derivative terms in eq.~\eqref{total_action_2} gives an irrelevant boundary term. The second condition implies that the operator $\R \, \delta K$, which contains $\partial^2 \psi \partial^2 \zeta$, does not appear. Finally, $\R^2 = 16 (\partial^2 \zeta)^2/a^4 $ and $ \R_{ij} \R^{ij} = [ 5 (\partial^2 \zeta)^2 + (\partial_i \partial_j \zeta)^2] /a^4 $: one can check that when the third condition is satisfied  the sum of  the two operators in eq.~\eqref{total_action_2} vanishes up to a total derivative.

In summary, the most general EFT Lagrangian which does not generate higher derivatives in the linear equations for the perturbations is\footnote{Let us comment here on the case of a non-vanishing spatial curvature, to which our formalism can be extended straightforwardly with the following caveats. Obviously, $\delta \, \R$ should be used instead of $\R$ in the quadratic operators, but apart from this the Lagrangian~\eqref{action_sec2} and its properties are unchanged. The background equations change (see e.g.~eqs.~16 and 17 or Ref.~\cite{GPV}), as well as the dictionary~\eqref{relations}, because some first order quantity will now contribute already at zeroth order. The explicit expressions~\eqref{RRR} of $\R$ change by a term linear in $\zeta$ but with no derivatives, which, therefore, will not produce higher derivatives in the ADM analysis of Sec~\ref{sec:ADM}. \label{footnote}} 
\be
\begin{split}
\label{action_sec2}
L &=  \frac{\MM^2}{2} f(t) R - \Lambda(t) - c(t) g^{00} + \, \frac{M_2^4(t)}{2} (\delta g^{00})^2\, -\, \frac{m_3^3(t)}{2} \, \delta K \delta g^{00} \, \\
  &-\,  m_4^2(t)\left(  \delta K^2- \d K^\mu_{ \ \nu} \, \d K^\nu_{ \ \mu}\right) \, +\, \frac{\tilde m_4^2(t)}{2} \, \R \, \delta g^{00} \;,
\end{split}
\ee
where 
\be
\label{definitions_ms}
m_3^3 \equiv \bar m_1^3 \;, \qquad m_4^2 \equiv \frac14 (\bar M_2^2 - \bar M_3^2) \;, \qquad \tilde m_4^2 \equiv \mu_1^2 \;,
\ee
as in eq.~\eqref{total_action}.
Terms containing $\R^2$ and $\R^{\,\mu}_{\, \nu }\, \R_{\,\mu }^{\,\nu}$ do not appear because they only contain higher spatial derivatives. 
By employing the dictionary \eqref{relations} in eq.~\eqref{zeta_Lagrangian}, the quadratic action for $\zeta$ is given by eq.~\eqref{quadratic_action_zeta}
where
\be
\label{zeta_Lagrangian_EFT}
\begin{split}
{\cal L}_{\dot \zeta \dot \zeta} &=  2 \left( c+ 2 M_2^4 -3 H^2 \MM^2 f - 3 H \MM^2 \dot f + 3 H m_3^3 - 6 H^2 m_4^2 \right){\cal D}^2+6 (\MM^2 f + 2 m_4^2) \;, \\
{\cal L}_{\partial_i \zeta \partial_i \zeta} &=  2 \left[  \MM^2 f - \frac{2}{a}\frac{d}{dt}(a\M) \right] \;,  
\end{split}
\ee
and 
\be
\begin{split}
{\cal D} & = \frac{2 (\MM^2 f +2 m_4^2)}{2 H (\MM^2 f +2 m_4^2) + \MM^2 \dot f - m_3^3 }\;, \\ 
\M&=  \frac{{\cal D}}2  ( \MM^2 f  + 2 \tilde m_4^2 ) \,.
\end{split}
\ee
The stability of a given model is then determined by the condition ${\cal L}_{\dot \zeta \dot \zeta} > 0$ and
the speed of sound can be straightforwardly computed from eq.~\eqref{sound_speed} by using the  relations above. One can check that these results agree with those found in~\cite{GPV} in the limit $m^2_4 = \tilde m^2_4 = 0$.

Finally, we can also write down the independent operators that generate higher spatial derivatives. These are 
\be
\begin{split}
L_{\rm h.s.d.} &=  - \, \bar m_4^2(t)\,  \delta K^2  \, +\, \frac{\bar m_5(t)}{2} \, \R \, \delta K + \frac{\bar \lambda(t)}{2}  \R^2  \, .
\end{split}
\ee

We now turn to study a well known example of  scalar-tensor theories of gravity which does not generate equations of motion with higher derivatives and, when restricting to linear perturbations, is contained in  {the Lagrangian} \eqref{action_sec2}.

\section{Generalised Galileons}
\label{section_Galileons}

\newcommand{\Gtwo}{G_2{}}
\newcommand{\Gthree}{G_3{}}
\newcommand{\Gfour}{G_4{}}
\newcommand{\Gfive}{G_5{}}
\newcommand{\Ftwo}{F_2{}}
\newcommand{\Fthree}{F_3{}}
\newcommand{\Ffour}{F_4{}}
\newcommand{\Ffive}{F_5{}}

In four dimensions, the most general scalar tensor theory having field equations of second order in derivatives is a combination of the following generalized Galileon Lagrangians~\cite{horndeski,Deffayet:2009mn,DGSZ},
\begin{align}
L_2 = & \Gtwo(\phi,X)\;,  \label{L2} \\
L_3 =& \Gthree(\phi, X) \Box \phi \;, \label{L3} \\
L_4 = &\Gfour(\phi,X)  R - 2 \Gfour_{X}(\phi,X) (\Box \phi^2 - \phi^{; \mu \nu} \phi_{; \mu \nu}) \;, \label{L4} \\
L_5 = & \Gfive(\phi,X)  G_{\mu \nu} \phi^{;\mu \nu} +\frac13  \Gfive_{X} (\phi,X)  (\Box \phi^3 - 3 \, \Box \phi \, \phi_{;\mu \nu}\phi^{;\mu \nu} + 2 \, \phi_{;\mu \nu}  \phi^{;\mu \sigma} \phi^{; \nu}_{\ ; \sigma}) \;. \label{L5}
\end{align}
For notational convenience, in this section we mostly indicate covariant differentiation with a semicolon symbol, i.e.~${}_{; \mu}$. Moreover, we have defined $X \equiv  \phi^{; \mu} \phi_{; \mu}$ (note that $X$ is sometimes defined differently, i.e., with a factor of $-1/2$).

In order to translate the above Lagrangians into our EFT language we will proceed in two steps.
First, we will rewrite each of these Lagrangians in terms of 3-dimensional geometrical objects ($K_{\ \mu}^{ \nu}$, $\R_{\ \mu}^{ \nu}$, etc.) so that their unitary gauge expression becomes easily readable.  The $3+1$ decomposition that we are after loses manifest general covariance but shows straightforwardly the lack of higher {\it time} derivatives already at the level of the action. The second step will be to compute the corresponding coefficients of the operators~\eqref{action_sec2} by simply inverting the dictionary that we derived in the previous section---eqs ~\eqref{barL} and \eqref{relations}:
\be
\begin{split}
c & = -\frac12\left(\dot \F+L_N\right) +  H \dot L_\sR - 2 L_\sR \dot H - \ddot L_{\sR}  \;, \\
\Lambda & =  - \bar L - \frac12 L_N + \frac12 \dot \F + 3 H \F + 2   \dot H  L_\sR +6 H^2 L_\sR + 5 H \dot L_\sR + \ddot L_\sR \;, \\ 
f & = 2 L_{\sR} \MM^{-2} \;, \\
M_2^4 &= \frac14 ( L_{NN} + 3 L_N  + \dot \F) - \frac12 ( H \dot L_\sR - 2 \dot H L_\sR - \ddot L_\sR )\;, \\
m_3^3 & = 2 \dot L_\sR - 2 H L_{\SK N} - L_{KN} =2 \dot L_\sR - \B \;,  \\
m_4^2 & = \frac12 \left( L_{\SK} - 2 L_\sR - 2H^2 L_{\SK \SK} - 2 H L_{\SK K} - \frac12 L_{KK}  \right) = \frac12 \left( L_{\SK} - 2 L_\sR \right) - \frac{1}{4}\A   \;, \\
\tilde m_4^2 & = L_{N \sR} \;, 
 \label{correspondence}
\end{split}
\ee
where we have directly adopted the notation \eqref{definitions_ms} which holds in absence of higher derivatives---more general relations are easily found when eq.~\eqref{dK_noder} is not satisfied.

The main result of this section is that the dynamics of linear perturbations for all generalized Galileons is described by~\eqref{action_sec2}, with the further restriction $m_4^2 = \tilde m_4^2$. This is in agreement with the result \cite{Deffayet:2009mn,DGSZ} that also higher {\it  space} derivatives are absent from the equations of motion. 
This section is rather technical; the reader uninterested in the details of the calculations  can skip the following 
subsections and go directly to Sec.~\ref{sec:summary} where we  summarize our main results.

\subsection{Geometric preliminaries}
In order to express in unitary gauge terms of increasing complexity, it is useful to review the geometric formalism
adapted to the $3+1$ decomposition and separate the quantities into ``orthogonal" and ``parallel" to the hypersurface $\phi = const$.  
First, we define the future directed unitary vector orthogonal to the hypersurface. Up to a factor $\gamma$, this is proportional to the gradient of $\phi$ , 
\begin{equation} 
n_\mu =- \gamma \, \phi_{; \mu}, \qquad \gamma =  \frac{1}{\sqrt{-X}}\, .
\end{equation}

The metric induced on the $\phi= const.$ hypersurface is $h_{\mu \nu} = n_\mu n_\nu + g_{\mu \nu}$. 
Orthogonal to $n_\mu$ are also various quantities that ``live" on the hypersurface, in the sense that they vanish when contracted with $n_\mu$:
the extrinsic curvature and the ``acceleration" vector
\begin{equation} \label{live}
 K_{\mu \nu} = h_\mu^\sigma\,  n_{\nu;\sigma}, \qquad \nd_\mu = n^\nu \, n_{\mu ; \nu} \, . 
\end{equation}
The last two equations can be inverted by decomposing the derivative of $n_\mu$ into 
parallel and parallel/orthogonal components, 
\begin{equation}
n_{\nu ; \mu} = K_{\mu \nu} - n_\mu \nd_\nu \, .
\end{equation}

By means of the quantities just defined, we can decompose the second derivative of the scalar field as 
\begin{equation}
\phi_{; \mu \nu} =- \gamma^{-1}(K_{\mu \nu} - n_\mu \nd_\nu - n_\nu \nd_\mu)+ \frac{\gamma^2}{2} \phi ^{; \lambda} X_{; \lambda} n_\mu n_\nu \label{phimunu}\, .
\end{equation}
Again, this decomposition into parallel and orthogonal quantities is useful when calculating complicated products such 
$\phi_{; \mu \nu}\, \phi^{; \nu \sigma} \, \phi_{\ \ ;\sigma}^{;\mu}$ that appear in $L_5$, see eq.~\eqref{L5}.

Other relevant equations  are the Gauss-Codazzi equations, relating the Ricci tensor and scalar intrinsic to the hypersurface, $\R_{\mu \nu}$ and $\R$, to the four-dimensional ones \cite{Poisson,Wald:1984rg},
\begin{align}
\R_{\mu \nu} &= (R_{\mu \nu})_\parallel + (n^\sigma n^\rho R_{\mu \sigma \nu \rho})_\parallel - K K_{\mu \nu} + K_{\mu \sigma} K^\sigma_{\ \nu}, \label{GC1}\\
\R &= R + K^2 - K_{\mu \nu}K^{\mu \nu} - 2 (K n^\mu  - \dot n^\mu)_{;\mu} \label{GC2} \, ,
\end{align}
where the symbol $\parallel$ means projection on the hypersurface of all tensor indices, e.g. $(V_\mu)_\parallel \equiv h^{\ \nu}_{\mu} V_\nu $.

\subsection{$L_3$}
Since $L_2$ is trivial, following \cite{GPV} we start from $L_3$ and see how to rewrite it in the EFT of dark energy formalism. First, it is convenient to define an auxiliary function $\Fthree(\phi,X)$ such that 
\be
\Gthree \equiv  \Fthree + 2 X \Fthree_{X} \;.
\ee
Thus, $L_3$ in eq.~\eqref{L3} can be written as
\be
L_3  =   \Fthree \Box \phi + 2 X \Fthree_{X} \Box \phi  \;.
\end{equation}

We integrate by parts the first term on the right-hand side and we rewrite the second term using $\Box \phi = -\gamma^{-1} K + \frac12 \phi^{; \mu} {X_{; \mu}}/{X}$, which is obtained by tracing eq.~\eqref{phimunu}. This yields
\be
L_3  = - ( \Fthree_{X} X_{;\mu} + \Fthree_{\phi} \, \phi_{; \mu})  \phi^{; \mu} - 2 X \gamma^{-1} \Fthree_{X} K +  \Fthree_{X}  {X_{;\mu}}\phi^{; \mu}   \;.
\ee
After noticing that the first term inside the parenthesis cancels with the last one we finally obtain an expression for $L_3$ which is of the form of eq.~\eqref{lag},
\be
\label{L3_new}
L_3  =   2   (- X)^{3/2} \Fthree_{X} K - X \Fthree_{\phi}  \;,
\ee
where we have used $\gamma = 1/\sqrt{-X}$.
In unitary gauge $\phi (t,\vec x)= \phi_0(t)$, which implies, for instance,
\be
\Fthree_{X} (\phi, X) \to \Fthree_{X} (\phi_0(t), - \dot \phi_0^2(t)/N^2) \;.
\ee
Using eq.~\eqref{correspondence},  it is now straightforward to derive the corresponding EFT parameters in terms of the Lagrangian parameters evaluated on the background.  They are explicitly given in App.~\ref{app:Galileon_dic} and coincide with those given in \cite{GPV}. 
They only depend on four Lagrangian parameters, $\Gthree_{\phi}$, $\Gthree_{X}$, $\Gthree_{X\phi}$ and $\Gthree_{XX}$, so that the dependence on the auxiliary function $\Fthree$ disappears. 
As expected from eq.~\eqref{L3_new},  $f$, $m_4^2$ and $\tilde m_4^2$ all vanish: in order to describe $L_3$ we only need $c$, $\Lambda$, $M_2^4$ and $m_3^3$.

\subsection{$L_4$}

We now proceed  with $L_4$, defined in eq.~\eqref{L4}.
Using eq.~\eqref{phimunu} and its trace we can rewrite this as
\be
\begin{split}
L_4 &= \Gfour R - 2 \Gfour_{X}\left[ \Big(\g^{-1} K + \frac{\g^2}{2} \phi^{; \mu} X_{; \mu} \Big)^2 - \g^{-2}(K_{\mu \nu}K^{\mu \nu} - 2 \nd_\mu \nd^\mu) - \frac{\gamma^4}{4} (\phi^{; \mu} X_{; \mu})^2\right]\\
&= \Gfour R + 2 X \Gfour_{X}(K^2 - K_{\mu \nu}K^{\mu \nu}) + 2 \Gfour_{X} X_{; \mu} (K n^\mu   - \nd^\mu )\;, \label{L4svil}
\end{split}
\ee
where in the second line we have used that $\gamma^{-2} = - X$. Moreover, for the last term we have replaced $\gamma^{-1} \phi^{; \mu}$ by $-n^\mu$ and  used $ \nd_\mu = \frac{\gamma^2}{2} h_\mu^{ \ \nu} X_{; \nu}$. In this last term we can employ that $\Gfour_{X} X_{;\mu} = \partial_\mu \Gfour - \Gfour_{\phi} \phi_\mu = \partial_\mu \Gfour + \gamma^{-1}\Gfour_{\phi} n_\mu $. After an integration  by parts this yields, using $n_\mu \dot n^\mu =0$,
\be
2 \Gfour_{X} X_{;\mu} (K n^\mu   - \nd^\mu ) = - 2 \Gfour   (K n^\mu - \dot n^\mu)_{; \mu} - 2 \gamma^{-1} \Gfour_{\phi}  K   \;.
\ee
The first term on the right-hand side of this expression can be rewritten by using the Gauss-Codazzi equation~\eqref{GC1}.
Plugging all this into the second line of eq.~\eqref{L4svil} we finally obtain $L_4$ in 3+1 decomposition, 
\begin{equation}
\label{L4_new}
L_4 = \Gfour {}^{(3)}\!R + (2 X \Gfour_{X} - \Gfour)(K^2 - K_{\mu \nu}K^{\mu \nu}) - 2 \sqrt{-X} \Gfour_{\phi}  K\, .
\end{equation}
It is now lengthy but straightforward to apply our usual dictionary \eqref{correspondence} to derive the corresponding EFT parameters.  Their explicit expression can be  found in App.~\ref{app:Galileon_dic}. They depend on the six Lagrangian parameters $G$, $\Gfour_{X}$, $\Gfour_{X\phi}$, $\Gfour_{XX}$, $\Gfour_{XX\phi}$ and $\Gfour_{XXX}$. We need all the seven parameters of the EFT action~\eqref{action_sec2} to describe $L_4$ but the last two are equal, $m_4^2 = \tilde m_4^2$.

\subsection{$L_5$}

This Galileon Lagrangian is more involved than the others.
Let us start working on the first term on the right-hand side of eq.~\eqref{L5}, $\Gfive \, G_{\mu \nu} \phi^{; \mu \nu}$. Integrating it by parts gives
\be \label{L5first}
\Gfive \, G_{\mu \nu} \phi^{; \mu \nu} = - \Gfive_{X} X^{; \nu} \phi^{; \mu} G_{\mu \nu} - \Gfive_{\phi} \gamma^{-2} G_{\mu \nu} n^\mu n^\nu \;.
\ee
At this stage, as we did for $L_3$, it is convenient to define an auxiliary function $\Ffive(\phi,X)$, such that 
\be
\Gfive_{X} \equiv   \Ffive_{X} + \frac{\Ffive}{2X}\;, \label{F5}
\ee
and use this definition to integrate by parts the term proportional to $\Gfive_{X}$ in~\eqref{L5first}. In particular, using that 
\be
\Gfive_{X} X_{; \rho}  = \gamma \nabla_\rho (\gamma^{-1} \Ffive) + \Ffive_{\phi} \gamma^{-1} n_\rho\;, 
\ee
we obtain
\be
\Gfive \, G_{\mu \nu} \phi^{; \mu \nu}= \Ffive  \phi^{; \mu \nu} G_{\mu \nu} +\gamma^{-2} ( \Ffive_{\phi} - \Gfive_{\phi}) G_{\mu \nu} n^\mu n^\nu -  \frac{\gamma}{2} {\Ffive} X^{; \mu} n^{\nu} G_{\mu \nu} \;. \label{1termL5}
\ee

Let us now work on the second term  on the right-hand side of eq.~\eqref{L5}.
Using eq.~\eqref{phimunu}  we can rewrite this as
\be
\frac13 \Gfive_{X} (\Box \phi^3 - 3\, \Box \phi \phi_{; \mu \nu}\phi^{; \mu \nu} + 2  \phi_{; \mu \nu}  \phi^{; \mu \sigma} \phi^{; \nu}_{\ ;\sigma}) = - \Gfive_{X} \frac{\gamma^{-3}}{3} {\cal K} +  \Gfive_{X} {\cal J}  \;, \label{2ndL5}
\ee
where
\begin{align}
{\cal K}& \equiv K^3 - 3 K K_{\mu \nu}K^{\mu \nu} + 2  K_{\mu \nu}  K^{\mu \sigma} K^\nu_{\ \sigma}\, ,\\
{\cal J}& \equiv - \frac12 \phi^{ ;\rho} X_{;\rho} (K^2- K_{\mu \nu}K^{\mu \nu}  ) - 2 \gamma^{-3}(K \nd_\mu \nd^\mu - K_{\mu \nu}\nd^\mu \nd^\nu)\, .
\end{align}
The term proportional to ${\cal J}$ on the right-hand side of eq.~\eqref{2ndL5} can be integrated by parts using the same trick as above, which yields
\be
\Gfive_{X}  {\cal J} =- F_{5} \g^{-1}\left( \frac12 {\cal K}  + K^{\mu \nu} n^\sigma n^\rho R_{\mu \sigma \nu \rho} - K n^\sigma n^\rho R_{\sigma \rho} + \nd^\sigma n^\rho R_{\sigma \rho}\right) - \frac{\gamma^{-2}}{2} \Ffive_{\phi} (K^2 - K_{\mu \nu} K^{\mu \nu}) \;. \label{G5XC}
\ee

For the last part of the calculation we need  the (one time-)contracted Gauss-Codazzi relation, eq.~\eqref{GC1},
which gives
\begin{equation}
K^{\mu \nu}G_{\mu \nu} = K^{\mu \nu} \R_{\mu \nu} - K^{\mu \nu}n^\sigma n^\rho R_{\mu \sigma \nu \rho} + K K_{\mu \nu}^2 -  K_{\mu \nu}^3 - \frac12 R K\, .
\end{equation}
Replacing $\phi^{; \mu \nu}$ with eq.~\eqref{phimunu} in eq.~\eqref{1termL5} and using this relation, the terms proportional to $\Ffive$ in eqs.~\eqref{1termL5} and \eqref{G5XC} combine and simplify to
\be
-\g^{-1} \Ffive \left(  \G_{\mu \nu} K^{\mu \nu} - \frac{1}{6} {\cal K} \right)\;.
\ee
Using this and putting together all the terms of $L_5$ from eqs.~\eqref{1termL5}, \eqref{2ndL5} and \eqref{G5XC}  we  finally obtain 
\begin{equation}
\begin{split}
L_5 = & \ - \sqrt{-X} \Ffive  \left( K^{\mu \nu} \R_{\mu \nu} - \frac12 K \R \right) - \frac{1}{3} (-X)^{3/2} \Gfive_{X} {\cal K}  \\
&+ \frac12 X (\Gfive_{\phi} - \Ffive_{\phi} ) {}^{(3)}\!R   + \frac{1}{2} X \Gfive_{\phi} (K^2 - K_{\mu \nu} K^{\mu \nu})\;, \label{L5final}
\end{split}
\end{equation}
where in the last line we have used
\be
2 G_{\mu \nu} n^\mu n^\nu = {}^{(3)}\!R + K^2 - K_{\mu \nu} K^{\mu \nu} \;.
\ee
Note that the last line of \eqref{L5final} has the same form as the first two terms of $L_4$ given in eq.~\eqref{L4_new}: by using eq.~\eqref{F5} it can be written as
\be
\Gfour_{}  {}^{(3)}\!R   + (2 X \Gfour_{X} - \Gfour) (K^2 - K_{\mu \nu} K^{\mu \nu})\;,
\ee
with $\Gfour \equiv \frac12 X (\Gfive_{\phi} - \Ffive_{\phi} )$.

In order to compute the coefficients of the various EFT operators we use the  dictionary \eqref{correspondence}. To treat the term $K^{\mu \nu} \R_{\mu \nu}$ we employ the prescription described by eq.~\eqref{lag_app2} in App.~\ref{app:KR}.
Moreover, it is useful to notice that, up to quadratic order, the combination $\cal{K}$ of the extrinsic curvature tensor can be replaced by an expression that depends only on $S$ and  $K$:
\beq
{\cal K}=6 H^3 - 6 H^2 K + 3 H K^2 - 3 H S + {\cal O}(3)\,.
\eeq
The EFT operator coefficients are explicitly given in App.~\ref{app:Galileon_dic}. One finds that they depend on the six Lagrangian parameters $\Gfive_\phi$, $\Gfive_X$, $\Gfive_{X \phi}$, $\Gfive_{X X}$, $\Gfive_{X X\phi}$ and $\Gfive_{X XX }$---the dependence on $\Ffive$ explicitly cancels out---and, as in the case of $L_4$, $m_4^2  = \tilde m_4^2$.
Thus, at linear order in the perturbations---quadratic in the action---$L_5$ does not bring any new operator with respect to $L_4$. The difference between $L_4$ and $L_5$ appears at the cubic order in the action.

\subsection{Summary} 
\label{sec:summary}

We have established a dictionary between  the generalized Galileon theory, eqs.~\eqref{L2}--\eqref{L5}, and the EFT of dark energy parameters entering the action \eqref{total_action}. Such a dictionary is explicitly given  in App.~\ref{app:Galileon_dic}. As summarised in Table~\ref{tab:Galileons}, the EFT operators and their associated time-dependent parameters that are needed to describe the generalized Galileons are only six: $c$, $\Lambda$ and $f$, the  three usual parameters  already present at the background level, and $M_2^4$, $m_3^3$, $m_4^2 = \tilde m_4^2$, progressively appearing in $L_2$, $L_3$, $L_4$ and $L_5$, contributing only to the perturbations. 
As already stressed, at quadratic order in the perturbations, $L_4$ contains the same number of independent operators as $L_5$---in particular, only the combination $m_4^2=\tilde m_4^2$ appears in the action. 
The case $m_4^2 \neq \tilde m_4^2$ encompasses  the generalized Galileons: it does not contain higher derivatives and yet does not belong to the generalized Galileon class.\footnote{Note that our formalism easily applies to nonlinear extensions of HordenskiÕs theories, such as described in Ref.~\cite{Appleby:2012rx}. In this particular case, one finds that $m^2_4\neq  \tilde m^2_4$ but the quadratic action contains higher order spatial derivatives.}  When $m_4^2 \neq \tilde m_4^2$, higher derivatives are expected to appear {\it beyond} linear order. However, the effect of these higher derivatives can be ignored as long as perturbations remain small and linear theory is a good approximation.

\begin{table}
\centering
  \begin{tabular}{|p{3cm}||c|c|c|c|c|c|}
 \hline
 & & &&&&\\[-2mm]
    \textbf{Operator} &
    $f$ &
    $\ \Lambda$ \ &
    $\ c\ $ &
    $\ M_2^4\ $ &
    $\ m_3^3\ $ &
    $\ m_4^2 =\tilde m_4^2$ 
    \\[2mm] \hline
       \hline
    $L_2$ & 0 & \checkmark & \checkmark & \checkmark & 0 & 0   \\ \hline
    $L_3$ & 0 & \checkmark & \checkmark & \checkmark & \checkmark & 0  \\ \hline
    $L_4$ & \checkmark & \checkmark & \checkmark & \checkmark &  \checkmark &  \checkmark  \\ \hline
    $L_5$ & \checkmark & \checkmark & \checkmark & \checkmark & \checkmark & \checkmark    \\ \hline
  \end{tabular}
  \normalsize
  \caption[List of operators associated with various models]{
 A list of the different contributions of the generalized Galileon Lagrangians \eqref{L2}--\eqref{L5} to the operators of~\eqref{action_sec2}.  }
\label{tab:Galileons}
\end{table}

\section{Observables}
\label{sec:obs}
\newcommand{\Fa}{\beta}
\newcommand{\psib}{\alpha}
\newcommand{\bmfs}{\bar m_4^2}
\newcommand{\bmfi}{\bar m_5}
\newcommand{\Mfs}{M_4^2}

Observables describing  large scale structures are computed in the framework of linear cosmological perturbation theory. In this section we first derive the perturbation equations describing the dynamics of dark energy and modified gravity. We include a matter sector describing cosmological species  such as cold dark matter, baryons, photons and neutrinos---by adding the matter Lagrangian ${\cal L}_m(g_{\mu \nu} , \psi_m)$ to eq.~\eqref{total_action}, so that the final Jordan frame action in unitary gauge reads
\be
\begin{split}
\label{final_action}
S =& \int d^4x \sqrt{-g} \Bigg[ \frac{\MM^2}{2} f(t) R - \Lambda(t) - c(t) g^{00} +\ \frac{M_2^4(t)}{2} (\delta g^{00})^2\, -\, \frac{m_3^3(t)}{2} \delta K \delta g^{00} \,  \\ 
&  -\,  m_4^2(t)\left( \delta K^2 - \delta K_{\mu \nu} \delta K^{\mu \nu} \right) + \frac{\tilde m_4^2(t)}{2} \R \, \delta g^{00}  \\ &- \bmfs (t) \delta K^2 + \frac{\bar m_5(t)}{2}  \R \, \delta K  + \frac{\bar \lambda (t)}{2} \R^2 + {\cal L}_m(g_{\mu \nu} , \psi_m)\Bigg] \;.
\end{split}
\ee
We  then discuss the modifications of gravity expected in linear theory.
We will use Newtonian gauge, which is often used in cosmology, especially to describe cosmological perturbations for modified gravity. Extension to other gauges or to so-called ``gauge invariant'' formalisms is straightforward.

\subsection{Perturbation equations}
\label{sec:pert_eqs}

We will first restore the general covariance of the action above and write it in a generic coordinate system. In order to do that we need to
reintroduce the scalar fluctuation $\pi$ via the Stueckelberg trick \cite{ghost,EFT1,EFT2}.
Under the time coordinate change $t \to t + \pi (t, \vec x)$, the four-Ricci scalar $R$ remains invariant, while  functions of time such as $f$ and the 3-dimensional quantities change as\footnote{With an abuse of notation, here we denote the extrinsic curvature on hypersurfaces of constant time  with $K_{ij}$ even when we are {\it not} in unitary gauge. The reader must be aware that $K_{ij}$ is not the same geometrical object {\it before} and {\it after} the Stueckelberg trick. The same also holds for $\R_{ij}$. In particular, after the Stueckelberg trick $K_{ij}$ and $\R_{ij}$ are respectively given by eq.~\eqref{KijRij}. }  \footnote{The operator $\tilde m_4^2$ is also considered in Ref. \cite{BFPW}. However, in v1 of this reference, the variation~\eqref{variationR} of $\R$ under the Stueckelberg trick has been overlooked 
and the error propagates into the Einstein equations and the various observables. With the authors of~\cite{BFPW} there is now agreement on this issue~\cite{private}.} 
\begin{align}
f &\to f + \dot f \pi  + \frac12 \ddot f \pi^2 \;,  \label{trans_ST_6} \\
g^{00} &\to g^{00} +  2 g^{0 \mu} \partial_\mu \pi + g^{\mu \nu} \partial_\mu \pi \partial_\nu \pi \;, \\
\delta K_{ij} &\to \delta K_{ij} - \dot H \pi h_{ij} - \partial_i \partial_j \pi  \;, \\
\delta K &\to \delta K  - 3 \dot H \pi - \frac1{a^2} \partial^2 \pi \;, \\ 
{}^{(3)}\!R_{ij} &\to {}^{(3)}\!R_{ij}  + H (\partial_i \partial_j \pi + \delta_{ij} \partial^2 \pi) \;, \\
{}^{(3)}\!R &\to {}^{(3)}\!R + \frac4{a^2} H \partial^2 \pi \;. \label{variationR}
\end{align}

In the new coordinates we consider a linearly perturbed FLRW metric with only scalar fluctuations,
\be
\label{metric_general}
ds^2 = - (1+2 \Phi) dt^2 + 2 \partial_i \psib \, dt dx^i + a^2(t)\left[ (1-2 \Psi) \delta_{ij} + 2 \chi_{ij} \right] dx^i dx^j \;,
\ee
where $\chi_{ij}$ is traceless and given in terms of the scalar perturbation $\Fa$, $\chi_{ij} \equiv ( \partial_i \partial_j - \frac13 \delta_{ij} \partial^2 ) \Fa $. The extrinsic curvature and the 3-dimensional Ricci tensor of the {\it new} equal-time hypersurfaces thus  read
\be
\begin{split}
\label{KijRij}
K_{ij} &=  e^{-\Phi} (H-\dot \Psi) h_{ij} + \dot \chi_{ij}- \partial_i \partial_j \psib  \;, \\
{}^{(3)}\!R_{ij} &= \partial_i \partial_j \Psi + \delta_{ij}  \partial^2 \Psi  + 2 \partial_k \partial_{(i} \chi_{j)}^{\ k}  - \partial^2 \chi_{ij}\;.
\end{split} 
\ee
We also decompose the matter stress-energy tensor at linear order as 
\begin{align}
T^0_{\ 0} &\equiv - (\rho_m + \delta \rho_m) \;, \label{se1}\\
T^0_{\ i} &\equiv (\rho_m + p_m) \partial_i v = - a^2 T^i_{\ 0}\;, \label{se2} \\
T^i_{\ j} &\equiv (p_m + \delta p_m) \delta^i_j + \left( \partial^i \partial_j - \frac13 \delta^i_j \partial^2 \right) \sigma \label{se3}\;,
\end{align}
where $\rho_m $ and $p_m$ are respectively the background energy density and pressure and $\delta \rho_m $ and $\delta p_m$ their perturbations, $v$ is the 3-velocity potential and $\sigma$ the scalar component of the anisotropic stress.
The background equations derived from the action \eqref{final_action} are \cite{GPV}
\begin{align}
c+ \Lambda &= 3 \MM^2 (f H^2 + \dot fH) -\rho_m \;, \label{background1} \\
\Lambda - c &= \MM^2 (2 f \dot H + 3 f H^2 + 2 \dot f H + \ddot f) + p_m \label{background2} \;.
\end{align}

Using these expressions and the transformations \eqref{trans_ST_6}-\eqref{variationR} allows us to rewrite  \eqref{final_action} as an action for the scalar fluctuations $\Phi$, $\psib$, $\Psi$, $\Fa$ and $\pi$. We can vary eq.~\eqref{final_action} expanded at second order and
then fix the Newtonian gauge by setting $\psib = 0 =\Fa$ in the equations derived. This yields five equations,
\begin{align}
\label{Hc}
0= \left. \frac{1}{\sqrt{-g}}\frac{\delta S}{\delta \Phi}\right|_{\psib = 0 =\Fa}  & \equiv A_{ \Phi} \Phi +  A_{\dot \Psi} \dot \Psi + A_{\pi} \pi + A_{\dot \pi} \dot \pi  + \frac{k^2}{a^2} ( A^{(2)}_{\Psi} \Psi  + A^{(2)}_{\pi} \pi  ) + \delta T^0_{\ 0}  \;, \\
\label{Mc}
 0= \left.\frac{1}{ \sqrt{-g} }\frac{\delta S}{\delta \psib}\right|_{\psib=0=\Fa} &\equiv k^2 \left[ B_{  \Phi} \Phi+ B_{\dot \Psi} \dot \Psi +   B_{ \pi} \pi  + B_{ \dot \pi} \dot \pi  +  \frac{k^2}{a^2} (B^{(2)}_{ \Psi} \Psi + B^{(2)}_{ \pi} \pi) \right] -  i k^i  \delta T^0_{\ i}  \;, \\
 0=\left. \frac{1}{ \sqrt{-g}}\frac{\delta S}{\delta \Psi} \right|_{\psib=0=\Fa} &\equiv C_{ \Phi} \Phi+ C_{\dot \Phi} \dot \Phi +  C_{\dot \Psi} \dot \Psi + C_{\ddot \Psi} \ddot \Psi + C_{\pi} \pi + C_{\dot \pi} \dot \pi + C_{\ddot \pi} \ddot \pi \nonumber \\ 
 &+ \frac{k^2}{a^2}(C^{(2)}_{ \Phi} \Phi +  C^{(2)}_{\Psi} \Psi + C^{(2)}_{\pi} \pi + C^{(2)}_{\dot \pi} \dot \pi  ) + \frac{k^4}{a^4} (C^{(4)}_{\Psi} \Psi + C^{(4)}_{\pi} \pi) -   \delta T^k_{\ k}\;, \\
 \label{TT}
0=\left. \frac{1}{\sqrt{-g}} \frac{\delta S}{\delta \Fa} \right|_{\psib =0 =\Fa} &\equiv  \left(k_i k^j - \frac13 \delta_i^{j} k^2 \right)  \bigg[ k^i k_j \Big(  D^{(2)}_{ \Phi} \Phi + D^{(2)}_{ \Psi} \Psi + D^{(2)}_{ \dot \Psi} \dot \Psi + D^{(2)}_{  \pi} \pi+ D^{(2)}_{ \dot \pi} \dot \pi \nonumber \\ &+ \frac{k^2}{a^2} ( D^{(4)}_{  \Psi} \Psi + D^{(4)}_{  \pi} \pi) \Big) -  \delta T^i_{\ j}  \bigg] \;, \\
0=\left. \frac{1}{\sqrt{-g}}\frac{\delta S}{\delta \pi} \right|_{\psib =0 =\Fa} &\equiv E_{  \Phi} \Phi+ E_{ \dot \Phi} \dot \Phi+  E_{ \Psi} \Psi +  E_{ \dot \Psi} \dot \Psi + E_{ \ddot \Psi} \ddot \Psi + E_{\pi} \pi + E_{\dot \pi} \dot \pi + E_{  \ddot \pi} \ddot \pi  \nonumber \\ & + \frac{k^2}{a^2} (E^{(2)}_{ \Phi} \Phi  +  E^{(2)}_{ \Psi} \Psi +  E^{(2)}_{ \dot \Psi} \dot \Psi + E^{(2)}_{ \pi} \pi) + \frac{k^4}{a^4} (E^{(4)}_{ \Psi} \Psi + E^{(4)}_{ \pi} \pi)
\label{sfe} \;.
\end{align}
The coefficients $A_a, B_a, C_a, D_a$ and $E_a$ of these equations are detailed in App.~\ref{app:coefficients}.

The explicit expressions of the above  coefficients given in the appendix contain also higher-derivative terms---those proportional to $\bmfs$, $\bmfi$ and $\bar \lambda$. To compare with the usual Einstein equations, here we  rewrite these equations by replacing the components of the stress-energy tensor $T^\mu_{\ \nu}$ with their expressions given in eqs.~\eqref{se1}--\eqref{se3}.  For simplicity, we set $\bmfs =\bmfi =\bar \lambda=0$. We obtain:
\\

{\bf  $\bullet$  ${00}$-component ($\delta S/\delta \Phi=0$):}
\be
\label{00EE}
\begin{split}
 & \MM^2  \bigg[  - 2 f \left(   \frac{k^2 }{a^2} \Psi + 3 H \dot \Psi + 3H^2 \Phi \right)  + \dot f \left(   \frac{k^2}{a^2} \pi + 3 H^2 \pi  - 3 H (\Phi-\dot \pi)  - 3 (\dot \Psi+ H \Phi)  \right) + 3 H \ddot f \pi \bigg] \\ & - (\dot c + \dot \Lambda) \pi  
 + (2c + 4   M_2^4 +3 H m_3^3) \left(\Phi -\dot \pi \right) +( m_3^3 -4 H \mfs) \left[-\frac{k^2}{a^2} \pi+3 (H \Phi +\pi \dot H+ \dot \Psi )\right] \\&- 4 \frac{k^2}{a^2} \tmfs (\Psi + H\pi) =\delta \rho _m  \;.
\end{split}
\ee

{\bf  $\bullet$ ${0i}$-component  ($\delta S/\delta \psib=0$):}
\beq
\label{0iEE}
\begin{split}
&\MM^2 \big[ (H  \dot f - \ddot f) \pi +\dot f  \left(\Phi -\dot \pi\right)+2 f (H \Phi +\dot \Psi ) \big] - 2 c \pi - m_3^3 \left(\Phi -\dot \pi \right)+4 \mfs (H \Phi + \dot \Psi+ \dot H\pi   ) \\
& =  - \left(p_m+\rho _m\right) v \;.
\end{split}
\eeq

{\bf  $\bullet$ ${ij}$-trace component  ($\delta S/\delta \Psi=0$):}
\be
\label{ijtrace}
\begin{split}
& \MM^2 \bigg\{ 
 2 f  \left[  -\frac13 \frac{k^2}{a^2} (\Phi -\Psi )  +  (3 H^2 + 2\dot H) \Phi     + H(\dot \Phi+ 3 \dot  \Psi)  +  \ddot \Psi  \right] \\
&+ \dot f \left[ -\frac23 \frac{k^2}{a^2}  \pi + 2H \Phi+ 2 H  ( \Phi - \dot \pi) - (3H^2 +2 \dot H) \pi    + 2 \dot \Psi  + \dot \Phi - \ddot \pi  \right]\\
& + \ddot f \left[ - 2 H \pi + 2 (\Phi  - \dot \pi) \right] - f^{(3)} \pi \bigg\}   + (\dot \Lambda - \dot c) \pi + 2c(\Phi  -\dot \pi)  \\
& -\frac43 \frac{k^2 }{a^2} \left[   \tmfs (\Phi  - \dot \pi) +   \left(  H \mfs + (m_4^2)^{\hbox{$\cdot$}} \right) \pi +  \mfs  \dot \pi \right]\\
& + 4 (\dot H \mfs)^{\hbox{$\cdot$}} \pi + 4 \mfs \dot H  \dot \pi  - \left[  (m_3^3)^{\hbox{$\cdot$}} + 3 H m_3^3\right] (\Phi - \dot \pi)  -m_3^3 (\dot \Phi - \ddot \pi)   \\
&+ 4  \left[ H (m_4^2)^{\hbox{$\cdot$}} + 3 H^2 \mfs + \dot H \mfs  \right] \Phi  +4 (m_4^2)^{\hbox{$\cdot$}} \dot \Psi+4 \mfs H (3 \dot H  \pi +\dot \Phi  +3 \dot \Psi ) + 4 \mfs \ddot \Psi 
= \delta p_m  \;. 
\end{split}
\ee

{\bf $\bullet$  ${ij}$-traceless component  ($\delta S/\delta \Fa=0$):}
\be
\label{ijtraceless}
\begin{split}
&  \MM^2 \left[ f (\Phi-\Psi) + \dot f \pi \right]+  2 \left[ \mfs \dot \pi + \mfs H \pi + ( m_4^2)^{\hbox{$\cdot$}} \pi \right]
+ 2 \tmfs (\Phi - \dot \pi)  = \sigma \;.
\end{split}
\ee

By combining eqs.~\eqref{Hc} and \eqref{Mc} we obtain the relativistic generalization of the Poisson equation,
\be
\label{Poisson}
F_{\Phi} \Phi +F_{\dot \Psi} \dot \Psi + F_{\pi} \pi + F_{\dot \pi} \dot \pi + \frac{k^2}{a^2} (F^{(2)}_{\Psi} \Psi + F^{(2)}_{\pi} \pi  ) = \delta \rho_m - 3 H(\rho_m + p_m) v \equiv \rho_m \Delta_m\;,
\ee 
which can be also written as:
\\

{\bf  $\bullet$ Generalized Poisson equation:}
\be
\begin{split}
& - \frac{k^2}{a^2} \left[ (2 f \MM^2 + 4 \tmfs) \Psi - (\dot f \MM^2 - m_3^3 + 4 H \mfs - 4H \tmfs  ) \pi \right]  +(6 \MM^2 H^2 \dot f - 6H c - \dot c -\dot \Lambda + 3 m_3^3 \dot H) \pi \\
&-(2c + 4 M_2^4) \dot \pi 
- (3 \MM^2 H \dot f - 2 c - 4 M_2^4 ) \Phi - 3 \MM^2 \dot f \dot \Psi + 3  m_3^3 (\dot \Psi + H\Phi) = \rho_m \Delta_m\;.
\end{split}
\ee

Note that when $m_4^2 \neq \tilde m_4^2$ some of the equations contain terms with higher derivates: for instance, the terms with $k^2 \dot \pi$ in eq.~\eqref{ijtrace}, fourth line, and those with $\dot \pi $ in eq.~\eqref{ijtraceless}. However, the scalar propagating degree of freedom  satisfies a second order equation. Indeed, one can use  eq.~\eqref{ijtraceless} to remove the higher derivative terms  from eq.~\eqref{ijtrace} and  derive a purely second order equation for $\Psi$. This is even clearer in unitary gauge, where higher derivative are explicitly absent---see analysis of Sec.~\eqref{sec:ADM}.

\subsection{Modification of gravity}
\label{sec:mod_grav}

In order to derive the effective Newton constant, $G_{\rm eff}$, we consider the quasi static approximation, i.e.~we neglect the time derivatives in the equations of motion and we neglect the anisotropic stress, $\sigma=0$ in eq.~(\ref{TT}). This is a good approximation for scales much smaller than the sound horizon scale, i.e.~for $k \gg aH/c_s$. 
For models with small  or vanishing sound speed (see e.g.~\cite{Creminelli:2009mu}) or on scales longer than the sound horizon, a consistent treatment which takes into account the time derivatives should be undertaken. 

In the quasi-static limit, $G_{\rm eff}$ is defined by
\be
-\frac{k^2}{a^2} \Phi \equiv 4 \pi G_{\rm eff}(t,k) \rho_m \Delta_m\; .
\ee
Following \cite{DeFelice:2011hq,Silvestri:2013ne}, in order to write the Poisson equation in this form we can use eqs.~(\ref{TT}), \eqref{sfe} and \eqref{Poisson}. For $c_s \sim {\cal O} (1)$, we can neglect $D_\Phi$, $D_\Psi$, $D_\pi$, $E_\Phi$, $E_\Psi$, $F_\Phi$ and $F_\pi$ from these equations and the effective Newton constant is thus given by 
\begin{equation}
4 \pi G_{\rm eff} = - [ {\cal M}^{-1}]_{13}\;, \qquad {\cal M} \ \equiv \ 
\begin{pmatrix}
D_\Phi^{(2)} & D_\Psi^{(2)} + D_\Psi^{(4)} (k/a)^2  & D_\pi^{(2)} + D_\pi^{(4)} (k/a)^2 \\
E_\Phi^{(2)} & E_\Psi^{(2)} + E_\Psi^{(4)} (k/a)^2  & E_\pi (k/a)^{-2}+ E_\pi^{(2)} + E_\pi^{(4)} (k/a)^2 \\
0 & F_\Psi^{(2)}   & F_\pi^{(2)} 
\end{pmatrix}  \;.
\end{equation}
We can write it  in a slightly more compact form as
\be
\label{Geff}
 4\pi G_{\rm eff}(k) = \frac{a_{-2} (k/a)^{-2} + a_0 +  a_2 (k/a)^2 + a_4 (k/a)^4}{  b_{-2} (k/a)^{-2} + b_0  + b_2 (k/a)^2 }\;,
\ee
where
\be
\begin{split}
a_{-2} & =  D^{(2)}_\Psi E_\pi   \;, \\
a_0 & =  D^{(2)}_\Psi E^{(2)}_\pi  -
    D^{(2)}_\pi E^{(2)}_\Psi  + D^{(4)}_\Psi E_\pi \;, \\
a_2 & =    D^{(2)}_\Psi E^{(4)}_\pi - D^{(4)}_\pi E^{(2)}_\Psi - D^{(2)}_\pi E^{(4)}_\Psi  + D^{(4)}_\Psi E^{(2)}_\pi \;, \\
a_4 & =  - D^{(4)}_\pi E^{(4)}_\Psi + D^{(4)}_\Psi E^{(4)}_\pi  \;, \\
b_{-2} & =    D^{(2)}_\Phi E_\pi F^{(2)}_\Psi  \;, \\
b_0 & = D^{(2)}_\Psi E^{(2)}_\Phi F^{(2)}_\pi - D^{(2)}_\Phi E^{(2)}_\Psi F^{(2)}_\pi -
    D^{(2)}_\pi E^{(2)}_\Phi F^{(2)}_\Psi + D^{(2)}_\Phi  E^{(2)}_\pi F^{(2)}_\Psi \;, \\
b_2 & =  - D^{(2)}_\Phi E^{(4)}_\Psi F^{(2)}_\pi -  D^{(4)}_\pi E^{(2)}_\Phi F^{(2)}_\Psi + 
   D^{(2)}_\Phi E^{(4)}_\pi F^{(2)}_\Psi  +   D^{(4)}_\Psi E^{(2)}_\Phi F^{(2)}_\pi \;.
\end{split}
\ee

Another quantity often used to parameterize deviations from General Relativity  is the  ratio between the gravitational potentials $\gamma\equiv \Psi/\Phi$, which is given by
\be
\gamma = \frac{[{\rm com}( {\cal M})]_{32}}{[{\rm com} ({\cal M})]_{31}}\;,
\ee
where ${\rm com} ( {\cal M})$ denotes the comatrix of ${\cal M}$. This
reads
\be
\label{gamma}
\gamma= \frac{ c_{-2} (k/a)^{-2} + c_0  + c_2 (k/a)^2 }{ a_{-2} (k/a)^{-2} + a_0  + a_2 (k/a)^2 + a_4 (k/a)^4} \, ,
\ee
with 
\begin{align}
 c_{-2}&=    - D^{(2)}_\Phi E_\pi   \;, \\
 c_0  &=  D^{(2)}_\pi  E^{(2)}_\Phi - D^{(2)}_\Phi  E^{(2)}_\pi    \;, \\
 c_2 & =  D^{(4)}_\pi E^{(2)}_\Phi  - D^{(2)}_\Phi E^{(4)}_\pi  \;.
 \end{align}
The expressions for $G_{\rm eff}$ and $\gamma$, eqs.~\eqref{Geff} and \eqref{gamma}, generalize those given for instance in \cite{DeFelice:2011hq} in absence of higher derivative operators, in which case $a_2 = a_4 = b_2 = c_2 =0$. When also $a_{-2}=b_{-2}=c_{-2}=0$ we recover the results of \cite{GPV}.
Finally, we note that the numerator of $G_{\rm eff}$ equals the denominator of $\gamma$, which confirms the results of Ref.~\cite{Silvestri:2013ne}\footnote{It simply follows from
$[ {\cal M}^{-1}]_{13} = (\det {\cal M})^{-1}{[{\rm com} ({\cal M})]_{31}} $.}.

\section{Conclusion}

In this paper we lay down the basic building blocks for a systematic phenomenological study of dark energy and its cosmological perturbations. Following~\cite{GPV}, our basic assumptions are that {\it  a)} dark energy/modified gravity brings in at most one scalar propagating degree of freedom and that {\it  b)} the weak equivalence principle is satisfied---there exists a metric tensor universally coupled to matter. ÊWe use the effective field theory formalism developed for inflation in~\cite{EFT1,EFT2}, that Êis based on an expansion in number of perturbations rather than in number of fields. Indeed, expanding the action in number of fields~\cite{PZW,BF} becomes unpractical Êevery time that the background field configuration undergoes a large excursion. On the opposite, the main advantage of the present (non-covariant) approach is that an expansion in number of perturbations can always be consistently truncated at the desired order of approximation, in virtue of the empirical fact that perturbations are small on the largest scales. Moreover, our formalism is ``ready to go", in the sense that there is no need of solving for the background equations first.
Apart from the three operators $f$, $c$ and $\Lambda$ responsible for the background evolution~\cite{GPV}, every new operator is at least quadratic in the perturbations: it does not affect the background and its dynamical effects can be studied independently.

In particular, we consider only operators that are at most quadratic in the number of per\-tur\-ba\-tions---those needed for the linearized equations of motion---and we single out a set of seven operators that bring up to two derivatives in the equations of motion. To achieve this result, in Sec.~\ref{sec_2} we  provide a systematic treatment of any Lagrangian that can be written in ADM form as a general function of extrinsic and intrinsic 3-dimensional curvature tensors and of the lapse function. This is already enough to avoid  higher time derivatives in the equations of motion. Then, in Sec.~\ref{sec_2.3} we  identify specific combinations of the EFT operators that are required to avoid higher-order  spatial derivatives.  Some operators can be re-expressed into other ones, thus simplifying the EFT Lagrangians up to quadratic  order. 

The entire Horndeski, or ``generalized Galileon", theory can be written in this formalism  (Sec.~\ref{section_Galileons}): a relevant amount of work has gone into re-expressing all the generalized Galileon Lagrangians in their ADM form and obtaining their EFT formulation. At linear order, Horndeski theories can be described by a  total of six operators: only three quadratic operators in addition to those---$f$, $c$ and $\Lambda$---accounting for the background (see eq.~\eqref{total_action} with $m_4^2 = \tilde m_4^2$). 
This seems a substantial simplification if compared to the full covariant treatment and well represents the power of the non-covariant EFT approach. The two Galileon Lagrangians $L_4$ and $L_5$, despite their scaring looks~\eqref{L4}-\eqref{L5}, are affordable at linear order in the perturbations with the addition of just one operator with respect to those needed for $L_3$. 

At linear order, Horndeski theory is {\it not} the most general scalar-tensor theory with second-order dynamics. Indeed, for $m_4^2 \neq \tilde m_4^2$ there exists another operator beyond Horndeski that in unitary gauge gives equations of motion limited to second order in time and space derivatives. In some gauges (for instance in Newtonian gauge, see Sec.~\ref{sec:pert_eqs}), this operator  generates higher derivatives in the equations of motion but one can show that the dynamics of the propagating degree of freedom remains second order.
[At linear order, there exists another operator beyond the Horndeski theory (for $m_4^2 \neq \tilde m_4^2$) that still gives equations of motion limited to second order in time and space derivatives.]
Finally, we analyze also some higher {\it  spatial} derivative operators, those in eq.~\eqref{hsd}. 

The time dependent coefficients of our seven plus three operators described by actions \eqref{total_action} and \eqref{hsd} 
remain to be constrained or measured by observations. Indeed, in Sec.~\ref{sec:pert_eqs}
we  provide  the set of linear perturbation equations in Newtonian gauge 
by varying these actions with respect to scalar metric and field fluctuations in a generic gauge. As an illustration, using these equations we compute the effective Newton constant in the quasi-static approximation and the ratio between the two gravitational potentials (Sec.~\ref{sec:mod_grav}). The computation of these "observables"  should be considered  as a first step towards  a more general and systematic study of the impact of dark energy on cosmological perturbations in order to fully exploit future observational data.

\section*{Acknowledgments}
Conversations and/or correspondence with Jolyon Bloomfield,  Giulia Gubitosi, Ignacy Sawicki, Lorenzo Sorbo, Shinji Tsujikawa and George Zahariade are gratefully acknowledged.  
D.L.~is partly supported by the ANR (Agence Nationale de la Recherche) grant STR-COSMO ANR-09-BLAN-0157-01. F.V.~is partially supported by the ANR {\it Chaire d'excellence} CMBsecond ANR-09-CEXC-004-01.

\appendix

\section{Lagrangian dependence on $\R_{\mu \nu}K^{\mu \nu}$}
\label{app:KR}

In this appendix we show how to treat a dependence on 
\be
\Y \equiv \R_{\mu \nu}K^{\mu \nu} 
\ee
in the unitary gauge Lagrangian.

Let us first show the relation
\be
\label{app_RK}
\int d^4 x \sqrt{-g}\, \lambda(t) {}^{(3)}\!R_{\mu\nu}K^{\mu\nu}  =  \int d^4 x \sqrt{-g} \left[ \frac{\lambda(t)}{2} {}^{(3)}\!R\; K + \frac{\dot \lambda(t)}{2 N} \; \R \right]\;,
\ee
or, equivalently, 
\be
\label{app_RK2}
\int d^4 x \sqrt{-g} \left[ \lambda(t) {}^{(3)}\!R_{\mu \nu}K^{\mu \nu}- \frac{\lambda(t)}{2} {}^{(3)}\!R\; K  -\frac{\dot{\lambda}(t)}{2 N} \R\right] = 0 \;,
\ee
up to some irrelevant boundary terms.
Since $K=\nabla_\mu n^\mu$, the last two terms in the above integral can be simplified via an integration by parts, so that the expression reduces  to
\be
\int d^4 x\sqrt{-g}\, \lambda(t)\, \left(\R_{\mu \nu}K^{\mu \nu}+ \frac{n^\mu}{2} \nabla_\mu \R\right)\, .
\ee
It remains to show that this can be written as the integral of a total derivative.

Using the explicit expressions for the extrinsic curvature in the ADM decomposition, eq.~\eqref{KijADM}, and $n^\mu=- N g^{0\mu} $, the above expression can be rewritten as
\be
\int d^4 x \sqrt{h}\lambda(t)\left[\frac12\left(h^{ik}h^{jl}\dot{h}_{kl}\R_{ij}+{}^{(3)}\!\dot{R}\right) - \nabla^i N^j \R_{ij} - \frac12 N^i \nabla_i \R\right]\, ,
\ee
where $\nabla_i$ is the covariant derivative with respect to the three-metric $h_{ij}$.
The second term can be integrated by parts and then vanishes when combined  with the last term, as a consequence  of the  Bianchi identity $\nabla^i \G_{ij} =0$. 
Finally, the term in parenthesis can be rewritten as
\be
h^{ik}h^{jl}\dot{h}_{kl}\; \R_{ij}  + {}^{(3)}\!\dot{R} = h^{ik}h^{jl}\dot{h}_{kl}\; \R_{ij}  + \dot h^{ij} \R_{ij} + h^{ij} {}^{(3)}\!\dot{R}_{ij} =  h^{ij} \; {}^{(3)}\!\dot{R}_{ij} \;.
\ee
and it is known that  the last expression can be reexpressed as the divergence of a three-vector, i.e. $ h^{ij} \; {}^{(3)}\!\dot{R}_{ij} =  \nabla_i J^i $ (the very same property is used to derive Einstein's equations from the Einstein-Hilbert action\footnote{See  for instance Eq.~(7.5.14) of Ref.~\cite{Wald:1984rg}.}). We have thus proved Eq.~\eqref{app_RK}.

Let us now assume that the Lagrangian $L$ introduced in Eq.~\eqref{start_point} also contains an explicit dependence on $\Y$. By noting that 
$\Y$ is already a perturbative quantity, i.e. vanishes in the background, and can be decomposed as 
\beq
 \Y= H \sR+ {}^{(3)}R_{\mu\nu}\delta K^{\mu\nu}\,,
\eeq
where the first term on the right hand side is a first (and higher) order quantity while the second term is only second order, one immediately finds that the expansion of the Lagrangian,  up to quadratic order, will yield the following extra terms with respect to the expression \eqref{lag} obtained in the main body:
\be
\label{lag_app}
L(N,\SK,K,\sR,\Y, \Z) \supset L_\Y  \Y + \left( L_{N \Y} \d N  +  L_{K \Y} \d K +  L_{\SK \Y} \d \SK +  L_{\sR \Y} \d \sR  \right) H \delta \sR+\frac12 L_{\Y\Y}H^2 \delta \sR^2\;.
\ee
The first term  can be expressed in terms  of $\sR$ and $K$ by using eq.~\eqref{app_RK} with $\lambda = L_\Y $.  Expanding  up to second order 
then yields
\be
L_\Y \,  \Y = \frac12 \left( \dot L_\Y + 3 H L_\Y \right) \d \sR + \frac12 \left( L_\Y \delta K - \dot L_\Y \delta N \right) \d \sR +{\cal O}(3) + {\rm boundary} \ {\rm terms}\;,
\ee
so that the expansion of the full Lagrangian  now reads
\be
\label{lag_app2}
\begin{split}
L(N,\SK,K,\sR,\Y, \Z)&= \bar{L}- \dot \F - 3H \F+L_N \, \d N+ \frac12 \left( 2 L_\sR+ \dot L_\Y + 3 H L_\Y \right) \d \sR 
\cr
&+ L_\SK\, \d K^\mu_\nu \d K^\nu_\mu+\left(2H^2 L_{\SK\SK}+2H L_{\SK K}+\frac12 L_{KK}\right)\d K^2 +\frac12 L_{NN}\d N^2\cr
&+\frac12 \left(L_{\sR\sR}+H^2L_{\Y\Y}+2HL_{\Y\sR}\right) \, \d \sR^2
+\left(2H L_{\SK N}+ L_{KN}\right)\d K \d N
\cr
&+\left(2H L_{\SK \sR}+ L_{K\sR}+HL_{K\Y}+2H^2 L_{\SK \Y}+ \frac12 L_\Y\right)\d K \d \sR\cr & +\left( L_{N\sR}+HL_{N\Y}- \frac12 \dot L_\Y \right)\, \d N \d \sR+{\cal O}(3)\,.
\end{split}
\ee
In summary, an explicit dependence of the action on $\Y$ can easily be included in our treatment, via  the following substitutions in Eq.~\eqref{lag},
\be
\begin{split}
L_\sR & \to L_\sR+ \frac12 \dot L_\Y + \frac{3}{2} H L_\Y \;, \\
L_{\sR\sR} & \to L_{\sR\sR}+H^2L_{\Y\Y}+2HL_{\Y\sR} \;, \\
L_{N\sR} & \to L_{N\sR}+HL_{N\Y}- \frac12 \dot L_\Y \;,\\
\C  & \to \C+HL_{K\Y}+2H^2 L_{\SK \Y}+ \frac12 L_\Y \;.
\end{split} 
\ee

\section{Tensor modes}
\label{app:tensor}

In this appendix we study the propagation of tensor modes in the action \eqref{total_action}. We consider the spatial metric \cite{malda}
\be
h_{ij} = a^2(t) e^{2 \zeta} \hat h_{ij} \; , \qquad \det \hat h =1\;, \qquad \hat h_{ij} = \delta_{ij} + \gamma_{ij} + \frac12 \gamma_{ik} \gamma_{kj} \;,
\ee
with $\gamma_{ij}$ traceless and divergence-free,  $\gamma_{ii}=0 = \partial_i \gamma_{ij}$. Since tensor modes decouple from scalars, we can simply replace this metric into the action 
\eqref{total_action}  by setting scalar perturbations to zero, which yields
\be
S_{\gamma}^{(2)} =\int d^4 x \, a^3 \frac{\MM^2 f }{8} \left[  \left(1+\frac{2m_4^2}{\MM^2 f}\right) \dot{\gamma}_{ij}^2 -\frac1{a^2}(\partial_k \gamma_{ij})^2 \right]
\;,
\ee
where we used that, up to integration by parts,
\be
  \R =-\frac{1}{4 a^2}(\partial_i\gamma_{kj})^2 \;, \qquad  K = 3 H \;, 
\ee
\be
\delta K_{ij}^2 = \frac14 \dot{\gamma}_{ij}^2 \;, \qquad K_{ij}K^{ij}-K^2 = - 6H^2 +\frac14 \dot{\gamma}_{ij}^2 \;, 
\ee
and  the Gauss-Codazzi relation \eqref{RR}.
Thus, for $m_4^2 \neq 0$ the speed of sound of gravity waves is different from the speed of light, 
\be
c_T^2 =  \left(1+\frac{2m_4^2}{\MM^2 f}\right)^{-1}\;,
\ee
which confirms \cite{Gao:2011qe,DeFelice:2011uc} in the case of generalised Galileon theories.

\section{EFT parameters for generalized Galileons}
\label{app:Galileon_dic}

Here we explicitly give the EFT of dark energy parameters in terms of the Lagrangian \eqref{lag}, for the generalized Galileon Lagrangians eqs.~\eqref{L3}--\eqref{L5}.
All quantities in the expressions below are calculated on the background.
\\

{\bf $\bullet$ $L_3$:}
\begin{align}
f &= 0\, , \\ 
 \qquad  \Lambda &= \dot \phi^2 (\ddot \phi + 3 H  \dot \phi) \Gthree_{X} \, ,
\\ 
c &= \dot \phi^2 (- \ddot \phi + 3 H  \dot \phi) \Gthree_{X} +  \dot \phi^2 \Gthree_{\phi}\, ,
\\ 
M_2^4 &= \frac{\dot \phi^2}{2} (\ddot \phi + 3 H  \dot \phi) \Gthree_{X} - 3 H \dot \phi^5 G_{3,XX} -   \frac{\dot \phi^4}{2} G_{3,X\phi} \,,
\\ 
m_3^3 &= 2 \dot \phi^3 \Gthree_{X} \,, \quad m_4^2 = \tilde m_4^2 = 0 \,.
\end{align}

{\bf $\bullet$ $L_4$:}
\begin{align}
\MM^2 f &= 2 \Gfour \;, 
\\
\Lambda&= \frac12 \dot {\tilde \F} +3H \dot X \Gfour_{X}-18 H^2 \Gfour_{X}\dot\phi^2+6H \Gfour_{X\phi}\dot\phi^3+12H^2 \Gfour_{XX}\dot\phi^4 \;,
\\
c&=-\frac12 \dot {\tilde \F} +3H \dot X \Gfour_{X}-6H^2 \Gfour_{X}\dot\phi^2+6H \Gfour_{X\phi}\dot\phi^3+12H^2 \Gfour_{XX}\dot\phi^4 \;,
\\
M_2^4&=\frac14 \dot {\tilde \F} -\frac32 H \dot X \Gfour_{X}+6H \Gfour_{X\phi}\dot\phi^3 
+18H^2 \Gfour_{XX}\dot\phi^4 
-6H \Gfour_{XX\phi}\dot\phi^5
-12H^2 \Gfour_{XXX}\dot\phi^6 \;,
\\
m_3^3&=
2 \dot X \Gfour_{X}-8H \Gfour_{X}\dot\phi^2+4 \Gfour_{X\phi}\dot\phi^3+16H \Gfour_{XX}\dot\phi^4 \;, \\
m_4^2&= \tilde m_4^2 = 2\Gfour_{X}\dot\phi^2\,,
\end{align}
with
\beq
\tilde \F\equiv 2 \MM^2  H f + \MM^2 \dot f + \F=2 \dot X \Gfour_{X}-8H \Gfour_{X}\dot\phi^2\,.
\eeq

{\bf $\bullet$ $L_5$:}
\begin{align}
\MM^2 f &=- {\Gfive}_\phi \dot \phi^2 +2{\Gfive}_X \dot \phi^2 \ddot\phi  \; , \\
c&=-\frac12 \dot {\tilde\F} + \frac32 \MM^2 H \dot f - 3 H^2  {\Gfive}_\phi \dot \phi^2 - 3 H^3 {\Gfive}_X \dot \phi^3 + 3 H^2   {\Gfive}_{X\phi} \dot \phi^4 + 2  H^3 {\Gfive}_{XX}  \dot \phi ^5 \;, \\
\Lambda&= \frac12 \dot {\tilde \F} + 3 \MM^2 H^2 f + \frac32 \MM^2 H \dot f - 6 H^2  {\Gfive}_\phi \dot \phi^2 - 7 H^3 {\Gfive}_X \dot \phi^3 + 3 H^2   {\Gfive}_{X\phi} \dot \phi^4 + 2  H^3 {\Gfive}_{XX}  \dot \phi ^5 \;,
\\
M_2^4&=\frac14 \dot {\tilde\F} -  \frac34 \MM^2 H \dot f - \frac32 H^3 {\Gfive}_X \dot \phi^3 + 6 H^2   {\Gfive}_{X\phi} \dot \phi^4 + 6  H^3 {\Gfive}_{XX}  \dot \phi ^5 - 3 H^2 {\Gfive}_{XX\phi}  \dot \phi ^6 - 2  H^3 {\Gfive}_{XXX}  \dot \phi ^7 \;, \\
m_3^3&= 
\MM^2 \dot f - 4 H  {\Gfive}_\phi \dot \phi^2 - 6 H^2 {\Gfive}_X \dot \phi^3 + 4 H  {\Gfive}_{X\phi} \dot \phi^4 + 4 H^2 {\Gfive}_{XX}  \dot \phi ^5 \;, 
\\
m_4^2&= \tilde m_4^2 =  {\Gfive}_\phi \dot \phi^2 + H {\Gfive}_{X}\dot\phi^3-\Gfive_X \dot\phi^2\ddot\phi\, ,
\end{align}
with
\beq
{\tilde \F}\equiv 2 \MM^2  H f + \MM^2 \dot f+\F= 2 \MM^2 f H + \MM^2 \dot f - 2 H \Gfive_{ \phi}\dot \phi^2 - 2 H^2 \Gfive_{ X} \dot \phi^3\;.
\eeq
%%%%

\section{Coefficients of the perturbation equations}
\label{app:coefficients}
In this appendix we define the coefficients appearing in eqs.~\eqref{Hc}--\eqref{sfe}. For convenience we have used $\Mfs$ defined as 
\be
\Mfs \equiv 2 m_4^2 + 3 \bar{m}_4^2\;.
\ee
Moreover, from the background equations \eqref{background1} and \eqref{background2} and using the background conservation equation for matter, $\dot \rho_m + 3 H (\rho_m + p_m)=0$, one obtains
\be
\dot{c} + \dot{\Lambda}=-6Hc+6 H^2 \MM^2 \dot f + 3\MM^2\dot{f} \dot H  \;.
\ee
We will make use of this relation to simplify some of the terms and eliminate the dependence with respect to $\dot \Lambda$ and $\ddot \Lambda$.

{\bf $\bullet$ Variation with respect to $\Phi$:}
\begin{align}
A_{ \Phi} &=  2 c +4 M_2^4-6 H \left[ f H \MM^2+ \MM^2 \dot f- m_3^3+  H \Mfs \right] \;, \label{APhi}\\
A_{ \dot \Psi} &=  -3 \left[2 H \left(f \MM^2+\Mfs \right)+\MM^2 \dot f -m_3^3 \right] \;, \\
A_{ \pi} &=  3 H^2 \MM^2 \dot f + 3 m_3^3   \dot H -\dot c-\dot \Lambda - 6\Mfs  H \dot H+3  \MM^2 \ddot f  \nonumber \\ 
& =   6H c- 3 (H^2+\dot H) \MM^2 \dot f +3  \MM^2 \ddot f + 3 m_3^3   \dot H  - 6\Mfs  H \dot H \;,  \\ 
A_{ \dot \pi} &=  -2 c-4 M_2^4-3 H(m_3^3-\MM^2 \dot f)\;, \\
A^{(2)}_{  \Psi} &= - 2 f \MM^2 + 6H \bmfi -4 \tmfs \;, \label{APsi} \\ 
A^{(2)}_{ \pi} &= \MM^2 \dot f   -m_3^3 +2 H \Mfs -4 H \tmfs  + 6 H^2 \bmfi  \;.
\end{align}

{\bf $\bullet$ Variation with respect to $\psib$:}
\begin{align}
B_{ \Phi} &=  -m_3^3+2 H \left(f \MM^2+\Mfs \right)+\MM^2 \dot f \;, \\
B_{ \dot \Psi} &= 2 \left(f \MM^2+\Mfs \right) \;, \\ 
B_{ \pi} &=  -2 c+ 2 \Mfs\dot H+\MM^2(H \dot f-\ddot f)  \;, \\
B_{ \dot \pi} &=  m_3^3-\MM^2 \dot f \;, \\
B^{(2)}_{ \Psi} & = -2  \bmfi \;, \\ 
B^{(2)}_{ \pi} &= -2   ( \bmfs + H \bmfi)  \;.
\end{align}

{\bf $\bullet$ Variation with respect to $\Psi$:}
\begin{align}
C_{ \Phi} &= 3  \left[2 c +2 (3 H^2 + \dot H ) \Mfs \right. \nonumber \\ & \left. +2 f  \MM^2(3 H^2+2 \dot H)  - ( m_3^3)^{\hbox{$\cdot$}} +H \left(-3 m_3^3+ 4 \MM^2 \dot f+ 2 (\Mfs)^{\hbox{$\cdot$}}  \right)+2 \MM^2 \ddot f\right] \;,  \\
C_{ \dot \Phi} &=  -3 m_3^3+6 H \left(f \MM^2+\Mfs \right)+3 \MM^2 \dot f \;,\\
C_{ \dot \Psi} &=  6 \left(3 f H \MM^2+3 H \Mfs +\MM^2 \dot f+ ( \Mfs)^{\hbox{$\cdot$}} \right) \;,\\
C_{ \ddot \Psi} &=  6 \left(f \MM^2+\Mfs \right) \;,\\
C_{ \pi} &=  -3  \left[\dot c-\dot \Lambda -2  \dot H ( \Mfs)^{\hbox{$\cdot$}} -2 (\ddot H + 3 H \dot H) \Mfs   +\MM^2(2 \dot f \dot H+f^{(3)} + 3 H^2 \dot f + 2 H \ddot f )\right] \nonumber \\
& =  -3  \left[2 \dot c+6Hc -2  \dot H ( \Mfs)^{\hbox{$\cdot$}} -2 (\ddot H + 3 H \dot H) \Mfs   +\MM^2 \left( f^{(3)} - (3 H^2+\dot H)  \dot f+ 2 H \ddot f \right)\right] \;,  \\
C_{ \dot \pi} &=    3\left(-2 c+3 H m_3^3-2 H \MM^2 \dot f+2 \Mfs \dot H  + ( m_3^3)^{\hbox{$\cdot$}}-2 \MM^2 \ddot f\right) \;,\\
C_{ \ddot \pi} &=  3(m_3^3-\MM^2 \dot f)\;, \\
C^{(2)}_{ \Phi} &=  - \left(2 f \MM^2-  6 H \bmfi+4 \tmfs\right) \;,  \\
C^{(2)}_{ \Psi} &=  2 f \MM^2- 6 H \bmfi - 6 \dot {\bar m}_5 \;, \\
C^{(2)}_{ \pi} &=   -2   \left( \MM^2 \dot f+ ( \Mfs)^{\hbox{$\cdot$}}  + H \Mfs +3 H^2 \bmfi  +3 H \dot {\bar m}_5 \right)  \;,\\
C^{(2)}_{ \dot \pi} &=  -2 (\Mfs -2 \tmfs  + 3 H \bmfi) \;,\\
C^{(4)}_{ \Psi} &= 16 \bar \lambda \;, \\
C^{(4)}_{ \pi} &=  -2  \bmfi   + 16 H \bar \lambda \;.
\end{align}

{\bf $\bullet$ Variation with respect to $\Fa$:}
\begin{align}
D^{(2)}_{  \Phi} &=    \MM^2 f + 2 \tmfs - 3 \bmfi H  \;, \\
D^{(2)}_{  \Psi} &=    - \MM^2 f  \;, \\
D^{(2)}_{  \dot \Psi} &=  -3 \bmfi  \;, \\
D^{(2)}_{ \pi} &=  \MM^2 \dot f + 2 \mfs H + 2 ( m_4^2)^{\hbox{$\cdot$}} - 3 \dot H \bmfi \;, \\
D^{(2)}_{ \dot \pi} &=  2 \mfs - 2 \tmfs  \;, \\
D^{(4)}_{ \Psi} &= -8 \bar{\lambda} \;, \\
D^{(4)}_{ \pi} &=  \bmfi - 8 H \bar \lambda \;.
\end{align}

{\bf $\bullet$ Variation with respect to $\pi$:}
\begin{align}
E_{  \Phi} &=  6 c H+\dot c+H^2(9 m_3^3-6 \MM^2 \dot f)   +3(2 m_3^3-\MM^2 \dot f) \dot H-\dot \Lambda+3 H\left[4 M_2^4- 2 \Mfs \dot H+ ( m_3^3)^{\hbox{$\cdot$}} \right]+4 ( M_2^4)^{\hbox{$\cdot$}} \nonumber  \\
& =  12 c H+2\dot c+3 m_3^3( 3 H^2 + 2 \dot H) - 6  \MM^2 \dot f (2 \dot H + H^2)  +3 H\left[4 M_2^4- 2 \Mfs \dot H+ ( m_3^3)^{\hbox{$\cdot$}} \right]+4 ( M_2^4)^{\hbox{$\cdot$}} \;,  \\
E_{  \dot \Phi} &=  2 c+4 M_2^4+3 H(m_3^3-\MM^2 \dot f) \;, \\
E_{  \Psi} &=   3  \left[6 c H+\dot c+\dot \Lambda-3 \MM^2 \dot f(2 H^2+\dot H)\right] =   0 \;,  \\
E_{  \dot \Psi} &=  3  \left[2 c+3 H m_3^3-4 H \MM^2 \dot f- 2 \Mfs \dot H+ ( m_3^3)^{\hbox{$\cdot$}}  \right] \;,\\
E_{  \ddot \Psi} &=  3(m_3^3-\MM^2 \dot f) \;,\\
E_{  \pi} &=  - \left[ 6 \Mfs  \dot H^2  -3  ( m_3^3)^{\hbox{$\cdot$}} \dot H   + 6 H \dot c  -9 H \dot H m_3^3 + \ddot c-3  \MM^2 \dot H \ddot f   -6 H^2  \MM^2 \ddot f -3  m_3^3 \ddot H+ \ddot \Lambda \right] \nonumber \\ 
&  =  - \left[ 6 \Mfs  \dot H^2  -3  ( m_3^3)^{\hbox{$\cdot$}} \dot H   + 6 \dot H  c + 3  \MM^2 ( \ddot H +4 H \dot H) \dot f    -9 H \dot H m_3^3 -3  m_3^3 \ddot H \right] \;,   \\ 
E_{  \dot \pi} &=  -2 \left[3 H \left(c+2 M_2^4\right)+\dot c+2 ( M_2^4)^{\hbox{$\cdot$}}\right] \;,\\
E_{  \ddot \pi} &=  -2 \left(c+2 M_2^ 4\right) \;, \\
E^{(2)}_{  \Phi} &=  - \left[m_3^3+H \left(- 2 \Mfs -6 H \bmfi+4 \tmfs\right)-\MM^2 \dot f\right] \;,  \\
E^{(2)}_{  \Psi} &=   -  2\left[2 H \tmfs+ \MM^2 \dot f - 3 \bmfi \dot H+2 ({ \tilde m}_4^2)^{\hbox{$\cdot$}} \right] \;,\\
E^{(2)}_{  \dot \Psi} &=  2 \Mfs + 6 H \bmfi-4 \tmfs  \;,\\
E^{(2)}_{  \pi} &=  - \left[ 2 c- 4 \Mfs \dot H + 4  \tmfs \dot H + ( m_3^3)^{\hbox{$\cdot$}} +  4 H^2 \tmfs + H m_3^3 -12  \bmfi H \dot H+4 H ({ \tilde m}_4^2)^{\hbox{$\cdot$}} \right] \;, \\ 
E^{(4)}_{  \Psi} &=  -  2 \bmfi +16 H \bar \lambda \;,\\
E^{(4)}_{  \pi} &=  -  2 (\bmfs + 2 H \bmfi)  + 16 H^2 \bar \lambda \;.
\end{align}

{\bf $\bullet$ Generalized Poisson equation:}
\begin{align}
F_{\Phi} &= - 3 \MM^2 H \dot f + 2 c + 4 M_2^4  + 3 H m_3^3 \;, \\ 
F_{\dot \Psi}& = - 3 \MM^2 \dot f  + 3  m_3^3 \;,\\
F_{\pi} &= 6 \MM^2 H^2 \dot f - 6H c - \dot c -\dot \Lambda + 3 m_3^3 \dot H \,  = -3\dot H (\MM^2  \dot f  -  m_3^3 ) \;,   \\
F_{\dot \pi}& = -(2c + 4 M_2^4)\;, \\
F^{(2)}_{\Psi} &= - (2 f \MM^2 + 4 \tmfs) \;, \\
F^{(2)}_{\pi} &= \dot f \MM^2 - m_3^3 + 4 H \mfs - 4H \tmfs \;.
\end{align}

\end{document}